\documentstyle[11pt,aaspp4,nick]{article}
\def\approxgt{\lower.2em\hbox{$\buildrel > \over \sim$}}
\def\approxlt{\lower.2em\hbox{$\buildrel < \over \sim$}}

\def\ion#1#2{\rm #1\,\sc #2}
\def\HI{{\ion{H}{i}}}

\def\GII{{\ion{He}{ii}}}

\def\dim#1{\mbox{\,#1}}
\def\lya{\hbox{Ly$\alpha$} }
\def\lyb{\hbox{Ly$\beta$} }
\def\eos{``effective equation of state''}
\def\eg{{\it e.g.\/}}
\def\ie{{\it i.e.\/}}
\def\etal{{\it et~al.\/}}
\def\cm2{\hbox{cm$^{-2}$}}
\def\eq{equation }
\def\eqs{equations }
\newcount\hours
\newcount\minutes
\def\SetTime{\hours=\time
\global\divide\hours by 60
\minutes=\hours
\multiply\minutes by 60
\advance\minutes by-\time
\global\multiply\minutes by-1 }
\SetTime
\def\now{\number\hours:\ifnum\minutes<10 0\fi\number\minutes}

\begin{document}

\pagestyle{myheadings}
\markright{DRAFT: \today\hfill}

\def\placefig#1{#1}

\slugcomment{Submitted to {\it The Astrophysical Journal}}
\title{The Evolution of the Effective Equation of State of the IGM}

\author{Massimo Ricotti, Nickolay Y.\ Gnedin,
and J.\ Michael Shull}
\affil{Center for Astrophysics and Space Astronomy, Department of
 Astrophysical and Planetary Sciences, University of Colorado, Campus Box
389, Boulder, CO 80309\\
ricotti,gnedin,mshull@casa.colorado.edu}

\begin{abstract}

We develop a method to extract the \eos\ of the intergalactic medium
from the Doppler ($b$) parameter distribution of the low-density \lya
forest. We test the method on numerical simulations and then apply it
to published observations of the \lya forest at redshifts $z\simeq 0$
to $4$.  We find that the effective equation of state is close to
isothermal at redshift $z \sim 3$, indicating that a second reheating
of the IGM took place at $z\sim3$.  This reheating can plausibly be
identified with the reionization of \GII\ observed to occur at
$z\sim3$.
\end{abstract}

\section{Introduction\label{sec:intro}}

It is widely believed that density and temperature fluctuations in the
intergalactic medium (IGM) are responsible for the \lya forest of
absorption lines observed in the spectra of QSOs. This hypothesis is
supported by the results of numerical simulations that are able to
reproduce accurately the statistical properties and main features of the
observed high-resolution spectra (cf., \markcite{wei99}Weinberg \etal\
1999 and references therein). At the same time, observations of the
high redshift \lya forest with Keck telescopes (\markcite{hui95}Hu
\etal\ 1995; \markcite{lu96}Lu \etal\ 1996; \markcite{kir97}Kirkman \&
Tytler 1997) and of the local \lya forest with the Hubble Space
Telescope (HST)
(\markcite{shu98}Shull, Penton, \& Stocke 1999; \markcite{pen99}Penton,
Stocke, \& Shull 1999) have sufficient spectral resolution and signal-to-noise
to permit a detailed study of the thermal properties of the IGM and
comparisons with the output of cosmological simulations.

In hierarchical dark matter dominated cosmological models, the formation of
the \lya forest depends on two loosely related pieces of information:
the evolution of the dark matter and the thermal history of the
IGM.\footnote{The lack of understanding of how star formation proceeds
prevents us from making a (in principle existing) direct connection
between the evolution of the dark matter and the evolution of the
ionizing radiation in the universe, the latter being the primary
source for the evolution of the thermal state of the IGM. The QSO
contribution to the ionizing radiation background that dominates up to
redshift $z \sim 4$ is, instead, better known.}
Recent
work by \markcite{cro98}Croft \etal\ (1999) in large part solved the
problem of recovering the power spectrum of the dark matter from the
\lya absorption spectra. Thus, the thermal history of the IGM remains
the only piece of information still unrecovered from the observational
data on the \lya forest. This paper complements the work of
\markcite{cro98}Croft \etal\ (1999) by attempting to recover the
thermal history of the IGM from the data.

It has been shown (\markcite{hg97}Hui \& Gnedin 1997) that in
sufficiently low density IGM (cosmic overdensity $\delta\la5-10$) the
gas temperature is tightly (to within a few percent) related to the
gas density, with a power-law relation of the form
$T=T_0(1+\delta)^{\gamma-1}$, which we will call an ``effective
equation of state'' because it appears to relate the density and the
temperature of the gas.\footnote{The true equation of state is, of
course, the ideal gas equation of state $P=nk_bT$.}  The thermal
history of the IGM can then be described with high precision by the time
evolution of the two parameters, $T_0$ and $\gamma$. In the most
simple physical model of the photoionized IGM adopted in all currently
existing numerical simulations, this evolution is determined by the
balance of adiabatic cooling due to the expansion of the universe and
photoionization heating of hydrogen and helium. However, it is also
possible that Compton heating from hard X-rays (\markcite{mad99}Madau
\& Efstathiou 1999), radiative transfer effects, and time-dependent
\GII\ reionization can be important processes affecting the thermal
history of the IGM (\markcite{ab99}Abel \& Haehnelt 1999;
\markcite{fer96}Ferrara \& Giallongo 1999).

The most readily available piece of observational data is the joint
distribution of the column densities and Doppler ($b$) parameters of
absorption lines. Since the column density of an absorption line
strongly correlates with the density of the gas in which the line
originates (\markcite{mir96}Miralda-Escud\'e \etal\ 1996;
\markcite{her96}Hernquist \etal\ 1996), we can use the column density
of a line as a measure of the gas density.  At the same time, the
width of the line contains information about the gas
temperature. Thus, we can recover the information about the evolution
of the \eos\ of the IGM from the $b-N$ distribution of the \lya
forest.  This was also noted independently by \markcite{sch99}Schaye
\etal\ (1999).

Undoubtedly, a better way to recover the \eos\ could be developed,
using the \HI\ absorption spectra rather than $b-N$ distributions. However,
this method is not yet developed. In addition, $b-N$ distributions are
readily available as published data, whereas the absorption
spectra are not. Thus, we proceed with analyzing $b-N$ distributions
with the understanding that our results could in principle be improved
upon by using the full information available in the absorption
spectra.

At low redshifts ($z\sim0$) the amount of the intergalactic gas that
is shock heated is thought to be larger than at higher redshifts
(\markcite{cen99}Cen \& Ostriker 1999). This, in principle, can induce
unrecovable systematic effects into our method. However,
\markcite{da99}Dav\'e \etal\ (1999) have shown that the \eos\ applies
down to redshift $z \simeq 0$ even if the fraction of gas that is
shock heated and that does not belong to the simple power-law relation
increases at low redshifts. A relation between the column density and
the overdensity of \lya clouds also holds at $z=0$. We can thus be
confident that we can measure the \eos\ of the IGM at low redshifts as
well. Even at $z=0$, there still exist a population of absorbers that
are not shock heated and which define the lower cutoff of the $b-N$
distribution (\markcite{dav97}Dav\'e \etal\ 1999). The resulting \lya
lines are distinct, narrow features in the HST spectra (Penton \etal\
1999); it is this population that we select by our method. In other
words, the approach we adopt automatically selects against the
shock-heated gas and thus is quite suitable for measuring the \eos\
even at low redshift.

The paper is organized in four sections. In \S~\ref{sec:method} we
describe our method to measure the \eos\ of the IGM. We study the
output of numerical simulations for several cosmological models in
order to understand how the thermal broadening of the lines affects the
$b-N$ distribution. In \S~\ref{sec:sim_results} we describe the
results of the method on the synthetic spectra created using the
simulation outputs, and in \S~\ref{sec:z-evol} we apply the method to
real observations of the \lya forest at $0 \approxlt z \approxlt
4$. In \S~\ref{sec:summary} we discuss the results and their
implications on understanding the thermal history of the IGM.

\section{Method\label{sec:method}}

We follow three logical steps to derive the \eos. First, we analyze
the result of several numerical simulations using the Voigt profile
fitting code AUTOVP (\markcite{dav97}Dav\'e \etal\ 1997) to build
synthetic $b-N$ distributions used to check the method. We consider
several distinct cosmological models with different values of the
ionizing intensity at 1 Ryd, $J_{21}$ (in units of $10^{-21}$ erg \cm2
s$^{-1}$ Hz$^{-1}$ sr$^{-1}$), and the \eos. We also analyze different
random realizations of the same model at the same or higher
resolution. We use the Hydro-PM approximation to model the low-density
IGM, which allows us to scan a large set of cosmological models and
random realizations with sufficient numerical resolution, a task, 
currently unachievable with the full hydrodynamic simulations. We
demonstrate that the systematic errors introduced by the HPM simulations
are completely negligible compared to other systematic and statistical
errors present in the method and the observational data.

We find that the thermal broadening $b_T=(13$ km s$^{-1})
\sqrt {T/10^4{\rm K}}$ of a line with peak overdensity $\delta$
(we use the notation $\rho=\overline \rho (1+\delta)$ where $\rho$ is
the baryon density and $\overline \rho$ is the mean baryon density)
corresponds to the value of $b=b_M$ at the maximum of the
$b-N$ distribution. Values of $b<b_T$ are possible because of the spread
of the $N-\rho$ distribution and because the Voigt profile
fitting routine introduces some lines that do not correspond to
density peaks. Another reason why the cutoff of the $b-N$ distribution
is not sharp is that errors on $b$ and $N$ smooth out the
distribution. This effect is negligible for the simulated \lya lines
because we pre-determine the continuum flux, zero flux levels and
higher signal to noise, but it is important for the observational data.

Second, we find an efficient way to locate the maximum of the $b-N$
distribution. In order to use all the information available, we
use the Maximum Likelihood Analysis (MLA) method to fit a
parametric two-dimensional function to the data points in the $b
\otimes N$ space. This allows us not only to locate the maximum of the
distribution but also to study its shape. We will see that this could
be a powerful way to extract information on the physics of
the low-density \lya clouds. We apply the MLA method to the simulated
\lya lines and demonstrate that the parameters $T_0$ and $\gamma$ of
the \eos\ lie inside the $1 \sigma$ confidence level of the values that
we measure for all the models and random realizations.

Third, we apply the MLA to real data sets at redshifts $z=4 \to 0$,
and we measure the location of the maximum and the shape of the
distribution. We account for the errors on $b$ and $N$ that, as
already mentioned, spread the real shape of the distribution. At this
point, we use the results of the simulations to interpret the $b$
values at the maximum of the distribution as due to thermal
broadening. This gives us the relation between $T$ and $N$ but not the
\eos. We still need to know the relationship between $N$ and $\rho$,
therefore we use the results we obtained from the simulations in
conjunction with the Density Peak Ansatz approximation to determine this
relationship (this is our second hypothesis based on the simulation
results). It turns out that we only need to know the column density of
the absorption lines with zero peak overdensity, $N_{\overline \rho}$,
to derive the \eos\ within the observational accuracy. 

%The measured
%evolution of the \eos indicate a second reheating of the IGM at $z
%\simeq 3$ and a temperature that reaches $T_0 \simeq $. We interpret
%this reheating as the \GII\ reionization observed to occur around
%exactly that redshift. As has been noted by \markcite{ab99}Abel
%\& Haehnelt (1999), radiative transfer
%effects can account for and enhancement of the temperature of a factor
%1.5-2 respect to the optically thin approximation. This perfectly match
%with our measured \eos. Finally we run the 'realistic' simulation of
%$\Lambda$CDM model with an \eos that resemble closely the measured one
%by including the effect of \GII\ reionization. We apply our method to
%the synthetic \lya forest using the Voigt fitting routine AUTOVP and
%the Density Peak Ansatz approximation and we demonstrate that we can
%recover the \eos. 

\subsection {Simulations\label{ssec:sim}}

We use an approximate simulation technique called HPM, developed by
\markcite{gh98}Gnedin \& Hui (1998), to model the fluctuations in the
low-density IGM ($\delta\la10$). In its essence, the HPM method uses
a simple Particle-Mesh (PM) solver modified to account for the effect
of gas pressure. The predefined \eos\ is then used to compute the gas
temperature and pressure at every point directly from the value of the
cosmic gas density at this point. Thus, there is no need to introduce
a special equation for the gas temperature as in a full hydrodynamic
solver. As a result, the HPM approximation is only about 25\% slower
(due to the overhead of computing the equation of state) than a simple
PM solver. It is substantially faster than a full hydrodynamic solver
(due to both fewer computations at each time step and fewer
time steps), while delivering results that are sufficiently accurate
for our purposes. 

\def\caphcbn{%
The $b-N$ distributions for the full hydrodynamic simulation ({\it crosses\/})
and the HPM simulation ({\it filled triangles\/}) of the same SCDM
cosmological model, the same initial conditions, and the same resolution.}
\placefig{
\begin{figure}
%\epsscale{0.50}
\insertfigure{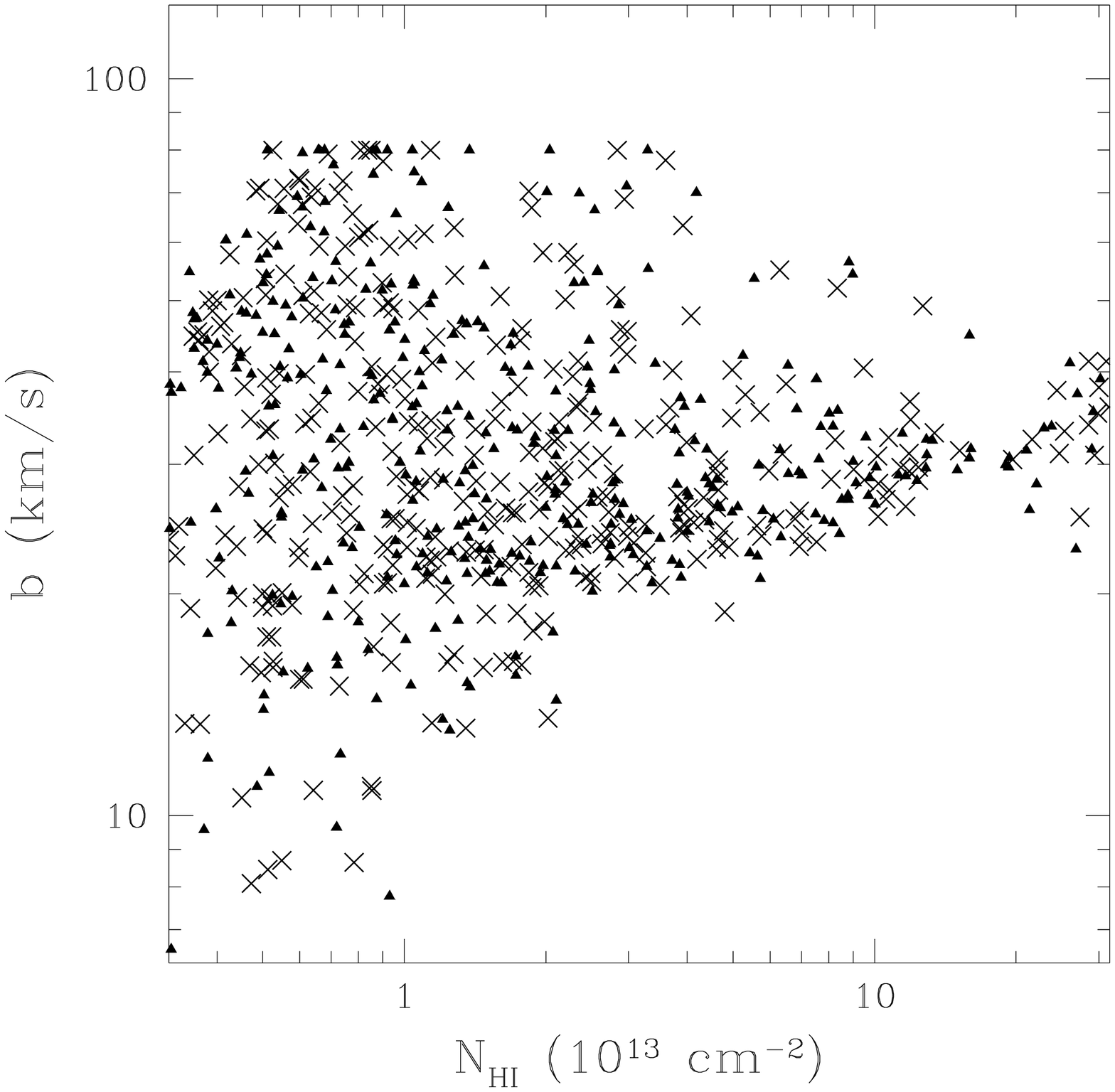}
\caption{\label{fig:hcbn}\caphcbn}
\end{figure}
}
In order to verify the accuracy of the HPM approximation, we performed
Voigt profile fitting for the full hydrodynamic simulations and for
the HPM simulation with the same initial conditions and resolution
used in the original HPM paper (Gnedin \& Hui 1998). 
% to about 15\%
%in the point-by-point comparison with a full hydrodynamic simulation
%of precisely the same cosmological model (for $\delta\la10$).
Figure
\ref{fig:hcbn} shows the $b-N$ distributions for the full hydrodynamic
simulation and the HPM approximation plotted on top of each other. 
The two distributions agree quite well.

\def\caphcbb{%
Comparison of $b$ parameters for the corresponding lines measured in the
HPM approximation versus the full hydrodynamic simulation.
The low-$b$ end of the distribution, where the maximum is located, is
matched better than the observational errors.}
\placefig{
\begin{figure}
%\epsscale{0.50}
\insertfigure{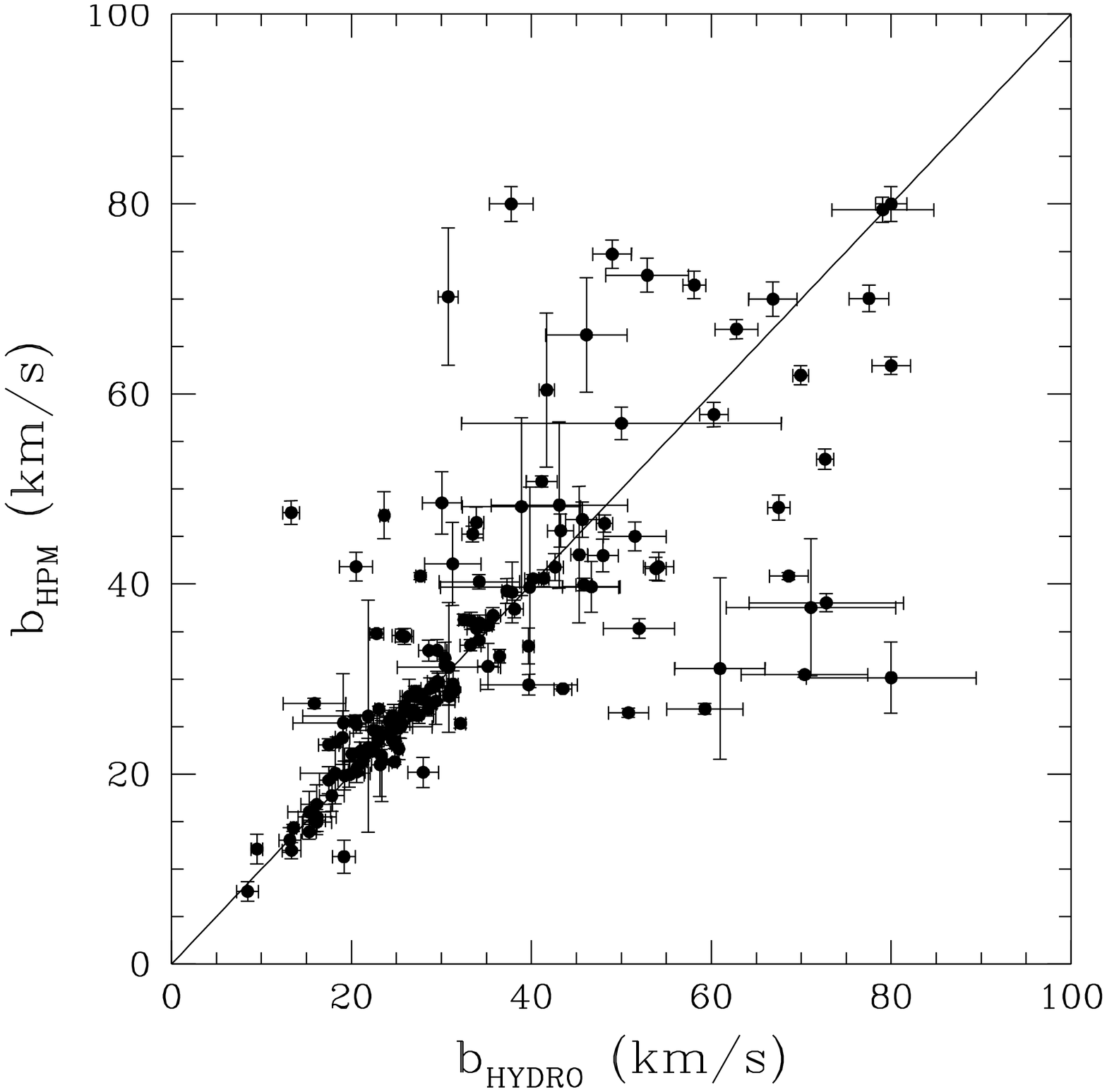}
\caption{\label{fig:hcbb}\caphcbb}
\end{figure}
}
Figure \ref{fig:hcbb} shows the comparison between the measured $b$-parameters
of the individual lines from two simulations. While at high $b$ values the
spread is significant, mostly due to the fact that the Doppler profile
fitting is not a uniquely defined procedure, at low $b$ values the
HPM and the full hydrodynamic simulation agree on a {\it line-by-line basis\/}
to better than observational
uncertainties. 

In order to estimate the systematic error in the position of the peak
of the $b-N$ distribution
introduced by the HPM simulation,
we have applied our method (described below) to the two distributions shown in
Fig.\ \ref{fig:hcbn}. We have found that
the intercepts, $b_0$, of the $b_T(N)$ relations (see \eq [\ref{eq:bN}]) 
agree within 0.5\%, and the slopes, $\beta$, agree within 3.5\%.
Thus, the systematic error in the position of the
introduced by the HPM simulation 
is completely negligible (at least a factor of 10 smaller)
compared to the other systematic
and statistical uncertainties (as discussed below in more detail).

%
% Table 1
%
\def\tabone{
\begin{deluxetable}{lccccccccccc}
\footnotesize
\tablecaption{Cosmological Models at $z=2.85$\label{tab:s1}}
\tablehead{
\colhead{Model} & \colhead{$\Omega_0$} & \colhead{$\Omega_b$} &
\colhead{$\Omega_\Lambda$} & \colhead{$\Omega_\nu$} & \colhead{$h$} &
\colhead{$\sigma_8$} & \colhead{$n$} & \colhead{$\gamma$} & 
\colhead{$T_{0,4}$} & \colhead{$\sigma_F$} & \colhead{$J_{21}$}}
\startdata
(a) SCDM.2A & 1    & 0.05  & 0    & 0    & 0.5  & 1 &  0.7  & 1.49 & 0.91 & 2.09 & 0.17\nl
(b) SCDM.2D$^{\rm a}$ & 1    & 0.05  & 0    & 0    & 0.5  & 1 &  0.7  & 1.49 & 0.91 & 2.09 & 0.17\nl
(c) SCDM.2E$^{\rm a}$ & 1    & 0.05  & 0    & 0    & 0.5  & 1 &  0.7  &
1.49 & 0.91 & 2.09 & 0.17\nl
(d) SCDM.L2A$^{\rm b}$ & 1    & 0.05  & 0    & 0    & 0.5  & 1 &  0.7  &
1.49 & 0.91 & 2.09 & 0.17\nl
(e) SCDM.2L & 1    & 0.05  & 0    & 0    & 0.5  & 1 &  0.7  & 1.33 & 1.21
& 2.26 & 0.17\nl
(f) LCDM.3A & 0.35 & 0.03  & 0.65 & 0    & 0.7  & 1 &  1.04 & 1.51 & 1.02
& 2.27 & 0.30\nl
(g) CHDM.3A & 1    & 0.05  & 0    & 0.2  & 0.5  & 1 &  0.81 & 1.49 & 0.91
& 1.41 & 0.25\nl
(h) OCDM.1A & 0.3 & 0.04 & 0   & 0    & 0.7  & 1 &  0.91 &
1.24 & 2.46 & 1.75 & 0.50\nl
 \enddata
\tablenotetext{a}{ A different random realization of the SCDM.2A
  model.}
\tablenotetext{b}{ $512^3$ cells and the box size of $5.12h^{-1}\dim{Mpc}$.}
\end{deluxetable}
}
\placefig{\tabone}
Numerical resolution is very important for accurately modeling the
\lya forest (Gnedin 1998; Theuns \etal\ 1998; Bryan \etal\ 1999).
The HPM simulations used in this paper have a cell size of
$10h^{-1}\dim{kpc}$ and resolve the filtering scale $k_F$ over which
the baryonic fluctuation is smoothed,
$$
        {\delta_b\over\delta_X} \approx e^{\displaystyle-k^2/k_F^2},
$$
(Gnedin \& Hui 1998) by about a factor of three. This choice of
the resolution scale has also been confirmed as sufficient by other authors
using the full hydrodynamic simulations 
(Theuns \etal\ 1998; Bryan \etal\ 1999). 

We produce synthetic spectra of the \lya forest at $z=2.85$ for eight of
the 25 different flat cosmological scenarios simulated using the HPM
approximation in \markcite{gne98}Gnedin (1998). The first three models,
SCDM.2A, SCDM.2D, and SCDM.2E, are three random realizations of the same
model, SCDM.2L is the same cosmological model with a different \eos,
LCDM.3A is a CDM model with the cosmological constant, CHDM.3A
represents a class of Cold+Hot CDM models, and OCDM.1A is an open CDM model. 
All of these simulations
have $256^3$ cells and thus have a box size of
$2.56h^{-1}\dim{Mpc}$.  In order to test the effect of box size,
we also analyze a larger simulation of the SCDM.2A model. This
simulation, labeled SCDM.L2A, has $512^3$ cells and box size of
$5.12h^{-1}\dim{Mpc}$.  As has been shown in Gnedin (1998), the
$5.12h^{-1}\dim{Mpc}$ size of the computational box is sufficient to
reduce the cosmic variance in the column density distribution to a few
percent.

The parameters of the models are summarized in Table~\ref{tab:s1}, where
we preserve the labeling used in \markcite{gne98}Gnedin (1998). We used
the AUTOVP Voigt profile fitting routine (\markcite{dav97}Dav\'e
\etal\ 1997) to measure $N$ and $b$ for absorption lines along a
thousand lines of sight in each simulation. This gives us about ten
thousand absorption lines per simulation. We use this $b-N$
distribution to verify the ability of our method to recover the
assumed \eos.

\subsection {Equation of State and $b-N$ Distribution\label{ssec:eos}}

\def\capd-N{% 
A scatter-plot (black dots) of the column density versus
density at the peak of density fluctuations for the model LCDM.3A at
$z=2.85$, calculated by using the Density Peak Ansatz approximation and
the simulations to measure the second derivative at the peak of the
density perturbations (see \S~\ref{ssec:n-d}). The long-dashed line
corresponds to the best fit using constant weighting of the lines and
the solid line using weighting $\propto N^{1/2}$ to account for the
0.5 slope of the logarithmic column density distribution.}
\placefig{
\begin{figure}
%\epsscale{0.50}
\insertfigure{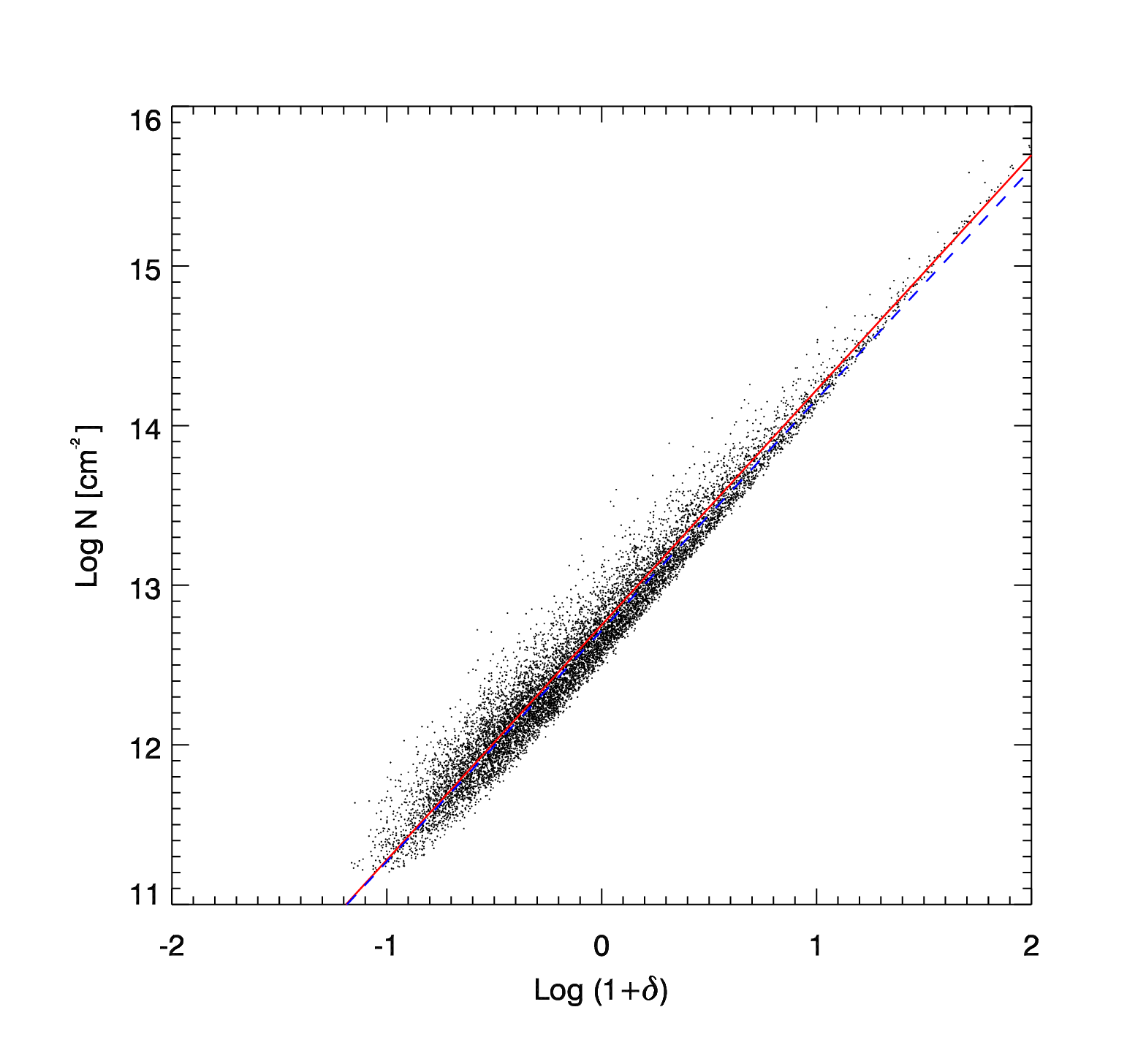}
\caption{\label{fig:d-N}\capd-N}
\end{figure}
}
There exists a tight correlation between the intergalactic gas density
and temperature in the low-density regime, where shock heating is not
important (\markcite{hg97}Hui \& Gnedin 1997). The density-temperature
relation is well described by a power-law \eos,
\begin {equation}
T=T_0 (1+\delta)^{\gamma-1},
\label{eq:Td}
\end {equation}
where $T_0$ is the temperature of the gas at the mean cosmic density
of baryons and $\gamma$ describes how the temperature changes with
density. It is clear that, if $\gamma=1$, the temperature of the IGM
is constant everywhere (the isothermal case). The evolution of both $T_0$
and $\gamma$ depends on the reionization history of the universe.
%but is not very
%sensitive to cosmological parameters. Assuming a sufficiently early
%reionization 
%$$
%T_0 \propto \left(\Omega_b h\over
%\Omega_0^{1/2}\right)^{0.6}
%$$
%(\markcite{hg97}Hui \& Gnedin 1997). If we
%trust the constraint on the baryon density $\Omega_b h^2=0.019 \pm 0.001$ from
%the measurements of the deuterium primordial abundance
%(\markcite{bt98}Burles \& Tytler 1998), then $T_0$ depends relatively weakly
%on cosmological parameters,
%$$
%       T_0 \propto \Omega_0^{-0.3} h^{-0.6}.
%$$

Thus, measuring the \eos\ would potentially allow us to uncover the evolution 
of the ionizing background in the universe, and $b-N$ distribution
of the \lya forest offers us a way to measure $T_0$ and $\gamma$ as
a function of redshift simply because the Doppler $b$ parameter is related
(albeit indirectly due to peculiar velocities) to the gas temperature,
whereas the column density of an absorber is known to be tightly correlated 
with the gas density.

\def\capb-N{% 
Distribution of $\log b$ versus $\log N$ for the SCDM.2D
model (black dots) at $z=2.85$. The contour plot is the $N-\rho$ distribution
re-mapped on the $b \otimes N$ space using the \eos\ and \eq
(\ref{eq:bT}). This corresponds to peculiar velocities being set to
zero. The solid line is the \eos\ 
rewritten in terms of the thermal width of the
lines as a function of the column density derived using \eq
(\ref{eq:bN}) to relate the column density and the gas density. As explained
in the text, we use only absorption lines in a small range of the column
densities. The range of the column densities used and the number of lines are
reported in Table \ref{tab:s2}.} 
\placefig{
\begin{figure}
\epsscale{0.70}
\insertfigure{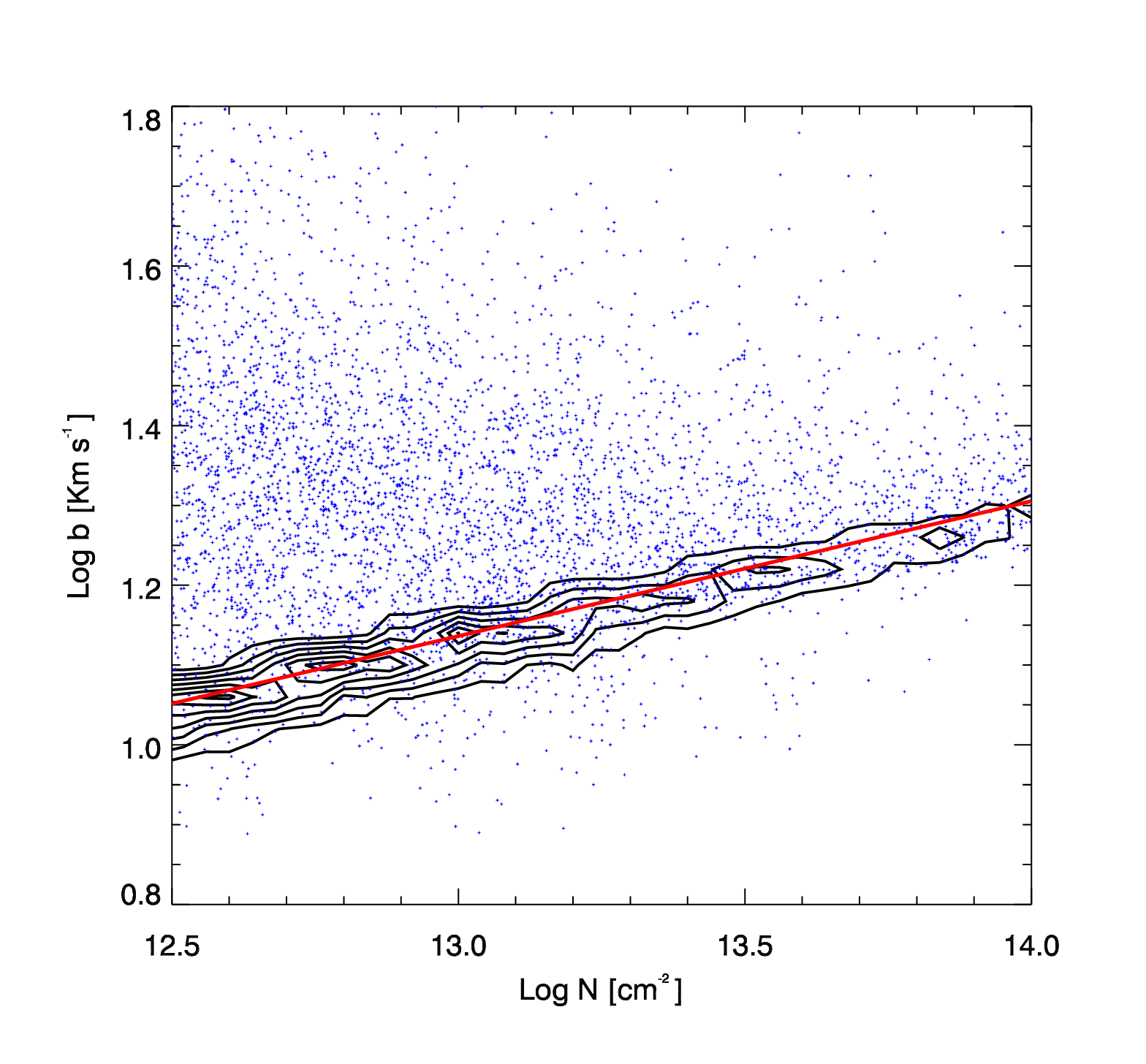}
\caption{\label{fig:b-N}\capb-N}
\end{figure}
}
To illustrate the latter point, we plot in Fig.~\ref{fig:d-N} 
the column density of the absorption
lines as a function of the overdensity at the peak of the line for the
LCDM.3A model at $z=2.85$ calculated by using the simulations together
with the Density Peak Ansatz
approximation (see \S~\ref{ssec:n-d}). In this case, a power-law,
\begin {equation}
N=N_{\overline \rho} (1+\delta)^\Gamma,
\label{eq:Nd}
\end {equation}
is also a good fit to the distribution, but the correlation between
$N$ and $\delta$ is not tight and the spread of the relation depends
on the cosmological model (\markcite{zan97}Zhang \etal\ 1996;
\markcite{hgz}Hui, Gnedin, \& Zhang 1997).

If we define the thermal broadening parameter,
\begin {equation}
b_T=\left({2kT \over m_H}\right)^{1\over2}=(13~{\rm km~s}^{-1})
T_4^{1 \over 2},
\label{eq:bT}
\end {equation}
where $T_4\equiv T/10^4 {\rm K}$, we can write the equation that
relates the thermal width and the column density of an absorption line
using (\ref{eq:Td}) and (\ref{eq:Nd}),
\begin {equation}
b_T=b_0 \left({N \over N_0}\right)^\beta,
\label{eq:bN}
\end {equation}
where,
\begin {eqnarray}
\beta&=&{\gamma-1 \over 2 \Gamma}, \\
b_0&=&(13~{\rm km~s}^{-1})T_{0,4}^{1/2}\left(N_0 \over N_{\overline \rho}\right)^\beta,
\end {eqnarray}
\def\capPb{% 
Histograms of a vertical slice at $z=2.85$ and $\overline N \simeq
10^{13}$ \cm2 of the scatter plot in Fig.~\ref{fig:b-N}. The solid
line represents the $b$-distribution with peculiar velocities being
set to zero (\ie, only thermal broadening $b=b_T$) and the long-dashed
line the $b$-distribution.}
\placefig{
\begin{figure}
%\epsscale{0.50}
\insertfigure{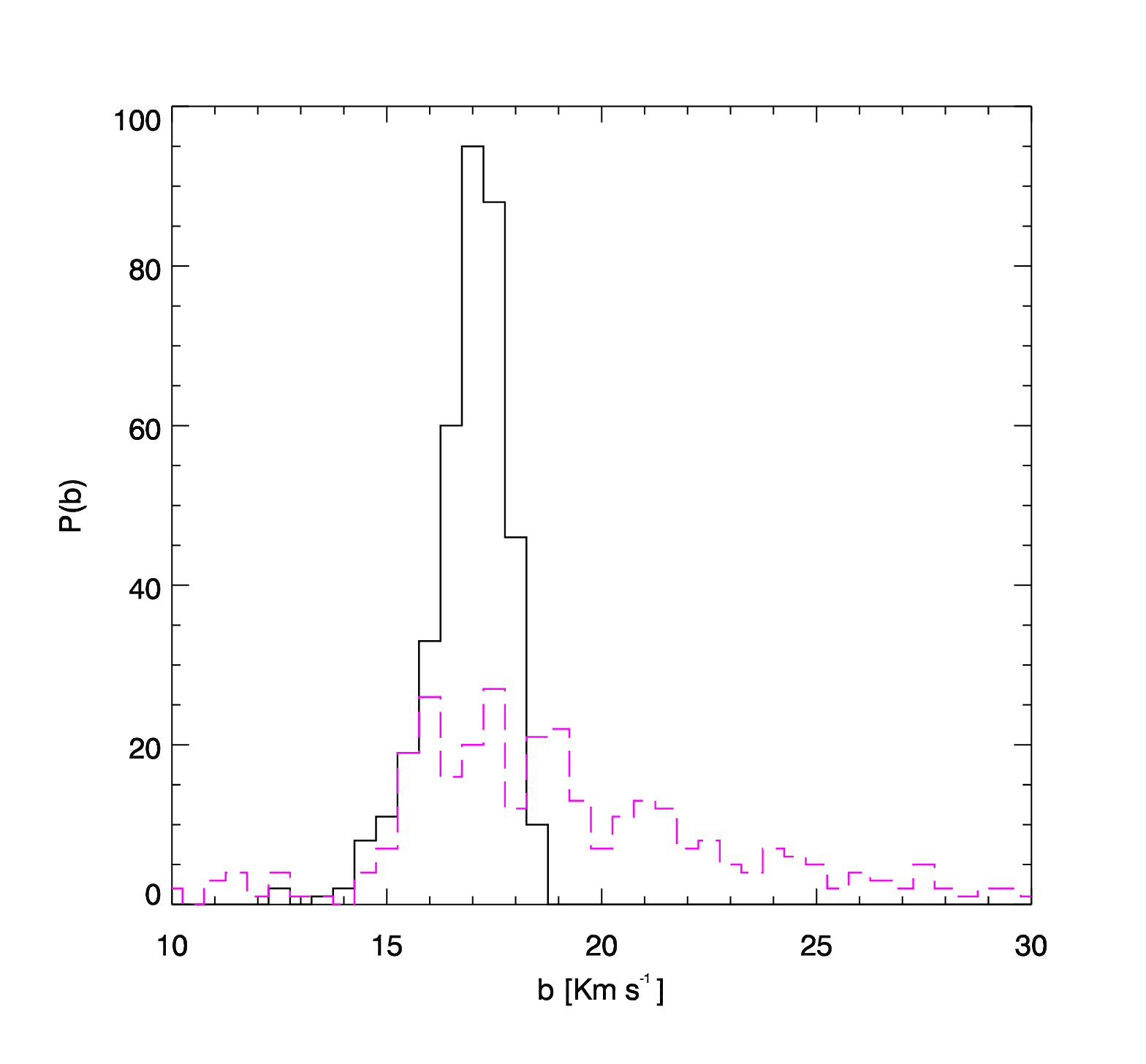}
\caption{\label{fig:Pb}\capPb}
\end{figure}
} 
where $T_{0,4}=T_0/10^4 {\rm K}$. For each simulation, the values of
$T_0$, $\gamma$, $N_{\overline \rho}$ and $\Gamma$ are known (the last
two are measured by fitting \eq (\ref{eq:Nd}) to the $N-\rho$ distribution as
shown in Fig.~\ref{fig:d-N} and as explained in
\S~\ref{ssec:n-d}). Therefore, using \eq (\ref{eq:bN}) we can derive
$b_T$. 
In Fig.~\ref{fig:b-N} we plot the $b-N$ distribution for 11,520
absorption lines (black dots), taken from the spectra of the 
SCDM.2D model with $J_{21}=0.17$ at redshift $z=2.8$ using AUTOVP
routine. The solid line is the thermal broadening $b_T$ defined in \eq
(\ref{eq:bT}), and the contour plot shows the $N-\rho$ distribution
remapped on the $b \otimes N$ space using the \eos\ and \eq
(\ref{eq:bT}), which corresponds to peculiar velocities being set to
zero.
It is evident that the solid line does not define the lower envelope
of the $b-N$ distribution owing to small-number statistics and to the intrinsic spread of the $N-\rho$
distribution. However, both the solid line and the lower envelope
track one another. Indeed, because we decided to fit the $N-\rho$
distribution instead of finding its upper cutoff (the cutoff toward
high column densities), the thermal broadening $b_T$ will correspond to
the peak of the $b-N$ distribution instead of its lower cutoff. 
In
Fig.~\ref{fig:Pb} we show this result more clearly by plotting a slice
of the distribution at the mean column density $\overline N \simeq
10^{13}$ \cm2, averaged on a bin of width $\Delta \log N \simeq
0.1$. The long-dashed line is the $b$-distribution and the solid line
is the $b$-distribution with peculiar velocities being set to
zero. Armed with this finding, in the next section we will seek an
efficient way to locate the maximum of the $b-N$ distribution and
demonstrate that, within the errors of the measurement, we are able to
recover $b_T$.

When we apply this method to the observed \lya forest in
\S~\ref{sec:z-evol}, we can only measure directly $b_T$ as a function
of the column density $N$. We do not
have a relation between the column density and overdensity in this
case. The parameters of the \eos\ are related to the measured $\beta$
and $T_{0,4}^\dag=(b_0/13~\rm{km~s}^{-1})^2$ by the equations:
\begin {eqnarray}
\gamma-1&=&2 \beta \Gamma, \label{eq:gamma} \\
T_{0,4}&=&T_{0,4}^\dag \left({N_{\overline \rho} \over N_0}\right)^{2 \beta}.
\label{eq:T0}
\end {eqnarray}
In \S~\ref{ssec:n-d} we find an approximate relationship for $\Gamma$
and $N_{\overline \rho}$ that allows us to solve the \eqs (\ref{eq:gamma}) and
(\ref{eq:T0}) given some cosmological parameters and the ionizing
background intensity $J_{21}$.

\subsection {Maximum Likelihood Analysis\label{ssec:mla}}

We define a likelihood function
\begin{equation}
{\cal L} = \prod_{i=1}^n ~p_i,
\end{equation}
where $p_i=p(b_i,N_i|model)$ is the probability for the line $i$ to
have a width $b=b_i$ and the column density $N=N_i$ and $n$ is the number
of lines in the sample. If we describe the magnitude of the errors by
a Gaussian distribution, then the probability function $p^{obs}_i$
observed is related to the true one by the double convolution:
\begin{equation} 
p^{obs}_i={(1-r^2)^{-{1\over2}} \over 2 \pi \sigma_{b,i}
\sigma_{N,i}}\int_{-\infty}^\infty \int_{-\infty}^\infty p(b,N|model)
\exp \left[-\left({\Delta b_i^2 \over 2 \sigma_{b,i}^2}+{\Delta N_i^2 \over
2 \sigma_{N,i}^2}-r{\Delta b_i \Delta N_i \over
\sigma_{b,i}\sigma_{N,i}}\right) {1 \over 1-r^2}\right]~dN~db
\end{equation}
where $\sigma_{b,i}$ and $\sigma_{N,i}$ are the errors on $b_i$ and
$N_i$ respectively, $\Delta b_i=b_i-b$, $\Delta N_i=N_i-N$ and $r$
 is the correlation coefficient  
between the errors in $b$ and $N$. 
%A positive value of r
%indicates that the errors are likely to have the same sign, while a
%negative value indicates that the errors are anticorrelated, likely to
%have opposite signs. 
Usually, the errors on the column density and Doppler
parameter tend to be anticorrelated ($r<0$), especially when the line is
saturated, since the fit tends to preserve the equivalent width of the
line. We adopt a parametric model for $p(b,N)$ and the free parameters
are determined by maximizing the likelihood function
(\markcite{efs88}Efstathiou \etal\ 1988).

A simple and accurate method of estimating errors is to determine
numerically the ellipsoid of parameter values defined by
\begin{equation}
\ln ~{\cal L} = \ln ~{\cal L}_{max} -{1 \over 2} \chi^2_\beta(M),
\end{equation}
where $\chi^2_\beta(M)=t \sqrt M$ is the $\beta$-point of the $\chi^2$
distribution with $M$ degrees of freedom and $t \sigma$ is the
confidence level wanted. The degrees of freedom are the number of
lines minus the free parameters of the distribution. The parametric
function, \eq (\ref{eq:pnb}), that we adopt to fit the $b-N$ distribution, has
nine free parameters, thus the ellipsoid of the errors is
nine-dimensional. We are only interested in determining the errors on $b_0$ and
$\beta$. The rigorous way to find these should be to project the
hyper-ellipsoid on the $b_0-\beta$ space but this is computationally
expensive. We use an approximate method that consists of
maximizing the likelihood function with respect to all the other
parameters for each value of $b_0$ and $\beta$.

In general, the matrix of errors is not diagonal, and the errors on the
parameters are correlated. The value of $N_0$ that makes errors
on $b$ and $\beta$ uncorrelated (\ie, the ellipse has principal axes
parallel to the Cartesian plane and the matrix of errors is
diagonal) is the average column density of the sample of
lines used in the fit. We choose $N_0$ in order to minimize the
correlation of the errors. The value of $N_0$ changes in each MLA
mainly because the range of the column densities of the data-set changes.

These are the main advantages of the method: (i) the MLA method has
well-defined asymptotic error properties (\eg, \markcite{ken61}Kendall
\& Stuart 1961); (ii) there is no requirement to select bin widths;
(iii) it is possible to account for errors on both $b$ and $N$; (iv)
we can also study the shape of the distribution; (v) all the
data are used; (vi) there is no need to find the
distribution cutoff that is not well defined because of artificial
lines and metal lines.
The disadvantage of this method is that it is difficult to test
whether the assumed parametric form of $p(b,N)$ is a good fit to the
data. In this paper, we adopt this method but we have also used a classic
minimum of Chi-square algorithm to fit the function $\overline
p(b)=p(b,N=\overline N)$ averaged over small intervals of the column
densities, selecting several bin widths of $N$. This preliminary study
has been useful to choose the form of the parametric function that is
a good fit to the data.

In the literature, the $b$ distribution is basically fitted with two
types of parametric functions: (i) a Gaussian with a low $b$ cut-off,
(\ie, a three-free-parameter function) (\markcite{kir97}Kirkman \&
Tyler 1997) and (ii) a one-parameter function of the form
(\markcite{hr99}Hui \& Rutledge 1999):
\begin {equation}
{dN \over db} \propto {b_\sigma^4 \over b^5} \exp \left[-{b_\sigma^4
\over b^4}\right],
\label{eq:n(b)}
\end {equation}
where $b_\sigma=[(b_T^2+b_{res}^2+b_J^2)/(D_+\sigma_0)]^{1/2}$ is
the $b$ value at the peak of the function. Here $b_J$ is the baryon
smoothing scale, $b_{res}$ is the resolution width, $D_+$ is the
linear growth factor ($D_+ = (z+1)^{-1}$ in a flat cosmology) and
$\sigma_0$ is the rms linear density fluctuation approximately at
the Jeans scale.  

We find that neither of these functions provides a good fit to the
data. Instead, we use \eq (\ref{eq:n(b)}) with the exponents of $b$ in
the power-law tail and in the exponential cutoff as free
parameters. We notice that, as the column density increases, the
power-law tail of the distribution becomes more narrow and the
exponential cutoff becomes more sharp. We have also tried a simple
power-law $b^{-n}$ with a low $b$ cutoff, motivated by the fact that
the exponential cutoff is very sharp, but we have found that this form
of the distribution is not a good fit to the data.

The form of the parametric function that is the best fit to the
simulated \lya forest and that we will use in the rest of the
paper is,
\begin {equation}
p(b,N)={\cal N} N^\alpha \left({b_* \over b}\right)^\varphi \exp
\left[{-\left({b_* \over b}\right)^\Phi}\right],
\label{eq:pnb}
\end {equation}

where,
\begin {eqnarray}
\varphi &=& \varphi_0 +\varphi_1 \left({N \over N_0}\right)+\varphi_2
\ln \left({N \over N_0}\right), \nonumber \\
\Phi &=& \Phi_0 +\Phi_1 \left({N \over N_0}\right)+\Phi_2 \ln \left({N
\over N_0}\right), \nonumber \\
b_*&=&b_M \left({\varphi \over \Phi}\right)^{1 \over \Phi}, \\
{\cal N} &=& \left({\alpha-1 \over
N_{min}^{1-\alpha}-N_{max}^{1-\alpha}}\right)\left({\Phi \over
b_*\Gamma \left[{\varphi-1 \over \Phi}\right]}\right).
\end {eqnarray}
Here, ${\cal N}$ is the normalization constant, $N_{min}$ and $N_{max}$ are
the low and high column density cutoffs of the distribution and
$\Gamma(z)=\int_0^\infty t^{z-1} e^{-t} dt$ is the Gamma function.
The $b$ value at the maximum of the function is $b_M$. As previously
stated, we assume $b_T=b_M$. We do not use the Hui \& Rutledge
relation for $b_\sigma$ because the Jeans smoothing scale in this
expression is needed only if we want to relate the dark matter density to
the baryon density and temperature. We already include this
smoothing scale in the $N(\rho)$ expression that we derive in
\S~\ref{ssec:n-d}. We assume that the broadening of the lines due to
the simulation/observation resolution is negligible in the range of
the column densities considered. We also find that the values of the
parameters $\Phi_1$,
$\Phi_2$ and $\varphi_2$ are practically zero in all the MLAs for the
simulations and observations as well. The shape of the distribution is
then determined by 6 parameters: $b_0$ and $\beta$ which determine the
value of $b_M$, $\varphi_0$ and $\varphi_1$ which are related to the
power law tail, $\Phi_0$ related to the exponential cutoff, and
$\alpha$ the slope of the column density distribution.

\def\capcor{% 
Value of the free parameters in \eq (\ref{eq:pnb}) as a function of the
correlation coefficient $r$. The parameters are normalized to their
value at $r=0$.}
\placefig{
\begin{figure}
%\epsscale{0.60}
\insertfigure{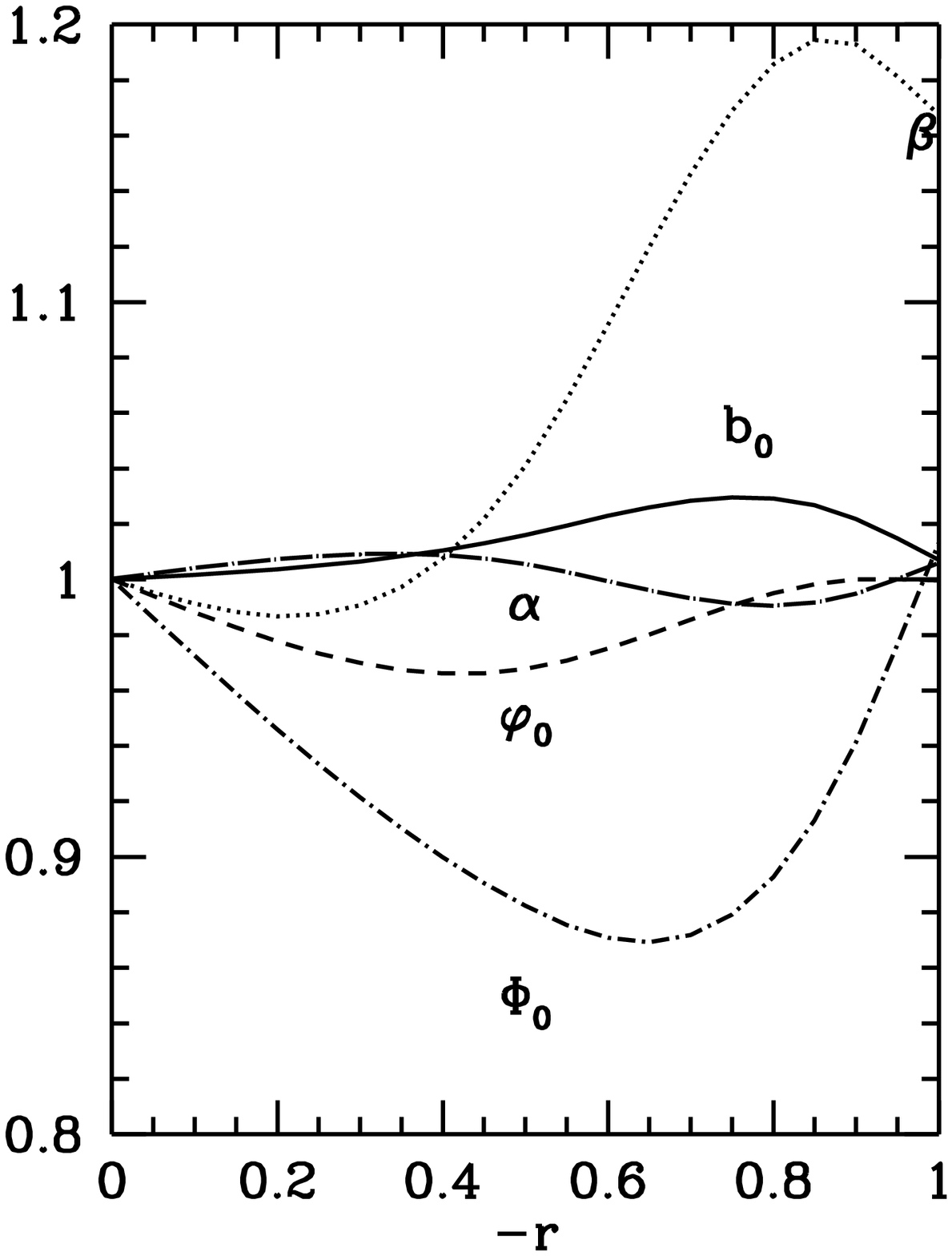}
\caption{\label{fig:cor}\capcor}
\end{figure}
} 
We do not know the correlation coefficient $r$ of the errors on $b_i$
and $N_i$. However, as is shown in Fig. \ref{fig:cor}, when we change
$r$ between zero and -0.95, the values of the parameters determined
with the MLA change by at most 20\%, well below the observational
accuracy or the accuracy of our method.

\section{Testing the Method \label{sec:sim_results}}

\subsection {The $b-N$ Distribution\label{ssec:b-n}}

The number of absorption lines in each simulation is about $10^4$,
but we are restricted to use lines in a range of the column
densities approximately one decade wide. This reduces the number of
lines to about $10^3$. The upper limit on $N$ is determined for each
simulation by imposing a maximum overdensity at the line center,
$\delta \sim 10$, which is the limit below which HPM simulations
produce accurate results.  The lower limit on the column density is
determined by the completeness requirement of the distribution
and by intrinsic problems associated with the Voigt fitting routine
AUTOVP. 
%In all of our simulations we observe a lack of lines in the
%region corresponding to underdense clouds with narrow b parameters. At
%a slightly larger value of the column density, corresponding to
%overdensity about $\delta=0$, we observe the lines with the bigger b
%values. It is possible that this is due to line blending but its
%origin is unclear.
The relative errors on $b$ and $N$ are about $10\%$, smaller than
typical observational errors, since we assumed a higher
signal-to-noise ratio than is usually afforded by the observational
data (in order to test our method in more challenging conditions) and
because we do not have uncertainties due to continuum fitting and zero
flux level. Therefore, the convolution with Gaussian errors does not
affect the shape of the distribution in a significant way.

%
% Table 2
%
\def\tabtwo{
\begin{deluxetable}{lcccccccccc}
\footnotesize
\tablecaption{Simulated \lya forest \label{tab:s2}}
\tablehead{
\colhead{Model} & \colhead{number} & \colhead{log $N_{min}$} &
\colhead{log $N_{max}$} & \colhead{log $N_0$} & \colhead{$b_0$} &
\colhead{$\beta$} & \colhead{$\varphi_0$} & \colhead{$\varphi_1$} & \colhead{$\Phi_0$} & \colhead{$\alpha$}\\[0.5ex]
\colhead{} & \colhead{of lines$^c$} & \colhead{(cm$^{-2}$)} & \colhead{(cm$^{-2}$)}
& \colhead{(cm$^{-2}$)} & \colhead{(km s$^{-1}$)} & \colhead{} & \colhead{} &
\colhead{} & \colhead{} & \colhead{}
 }
 \startdata
(a) SCDM.2A & 11500/1716 & 13.2 & 14.6 & 13.7 & 16.8 & 0.18 & 3.1 & 3.5 & 3.6 & 1.9 \nl
(b) SCDM.2D$^{\rm a}$ & 11520/1744 & 13.2 & 14.6 & 13.7 & 18.0 & 0.16 & 3.9 & 3.7 & 3.8 & 1.8 \nl
(c) SCDM.2E$^{\rm a}$ & 10222/1204 & 13.2 & 14.6 & 13.7 & 18.5 & 0.15 & 4.0 & 3.7 & 3.7 & 1.9 \nl
(d) SCDML.2A$^{\rm a,b}$ & 12131/1373 & 13.3 & 14.5 & 13.7 & 19.0 &
0.16 & 4.1 & 3.1 & 2.5 & 1.8 \nl
(e) SCDM.2L & 6980/1370 & 13.0 & 14.9 & 13.7 & 19.2 & 0.11 & 6.9 & 3.7 & 2.2 & 1.7 \nl
(f) LCDM.3A & 9233/1182 & 13.2 & 14.4 & 13.7 & 19.4 & 0.14 & 3.6 & 3.7 & 4.9 & 1.7 \nl
(g) CHDM.3A & 11812/1496 & 13.2 & 14.5 & 13.7 & 17.6 & 0.17 & 2.8 &
3.7 & 4.2 & 1.9 \nl
(h) OCDM.1A & 4289/271 & 13.5 & 14.9 & 13.7 & 26.1 & 0.04 & 12 & 0.1 &
13 & 1.9 \nl
% LCDM.6A & 4289/299 & 13.5 & 14.5 & 13.7 & 20.0 & 0.14 & 12 & 0.1 &
%13 & 1.9 \nl
% LCDM.6A & 4289/311 & 13.5 & 14.5 & 13.7 & 19.1 & 0.17 & 12 & 0.1 &
%13 & 1.9 \nl
% LCDM.6A & 4289/240 & 13.5 & 14.9 & 13.7 & 25.4 & 0.05 & 12 & 0.1 &
%13 & 1.9 \nl
% LCDM.6A & 4289/121 & 13.5 & 14.9 & 13.7 & 29.8 & 0.06 & 12 & 0.1 &
%13 & 1.9 \nl

 \enddata
\tablenotetext{a}{ A different random realization of the SCDM.2A
  model.}
\tablenotetext{b}{ $512^3$ cells and the box size of $5.12h^{-1}\dim{Mpc}$.}
\tablenotetext{c}{ Total number of lines/number of lines in the
column density range $N_{min} \le N \le N_{max}$.}
\end{deluxetable}
}
\placefig{\tabtwo}
In Table~\ref{tab:s2} we show the values of the parameters that give
the best fit to the $b-N$ distribution. 
We note that the value of
$\varphi_1$ is surprisingly constant for all the models but one. 
Also, the lack
of lines in the region of large $N$ and $b$ turns out to be a real
feature of the distribution and not an apparent feature due to a
decreasing number of lines. The width of the $b$ distribution at fixed
$N$ is proportional to $1/\varphi$ and decreases with $N$ as
$1/(\varphi_0+\varphi_1 N)$.
The value of $\Phi$, instead, varies between four and six and is directly
related to the width of the $N-\rho$ distribution.

%The parameter $\alpha$ is slightly bigger than the value $\alpha
%\simeq 1.5$ measured by many authors but our results are a coarse
%evaluation of the shape of the distribution and we do not determine
%the error of this parameter.

\def\capsm{% 
Error ellipses for the models at $z=2.85$, listed in
Table \ref{tab:s1}. The solid lines show $1\sigma$ error contours and the
dashed lines show the $2\sigma$ error contours. The crosses
are the expected values of $b_0$ and $\beta$ calculated
with \eq (\ref{eq:bT}). The size of the crosses corresponds to the
systematic error introduced by the HPM approximation.}
\placefig{
\begin{figure}
%\epsscale{0.70}
\insertfigure{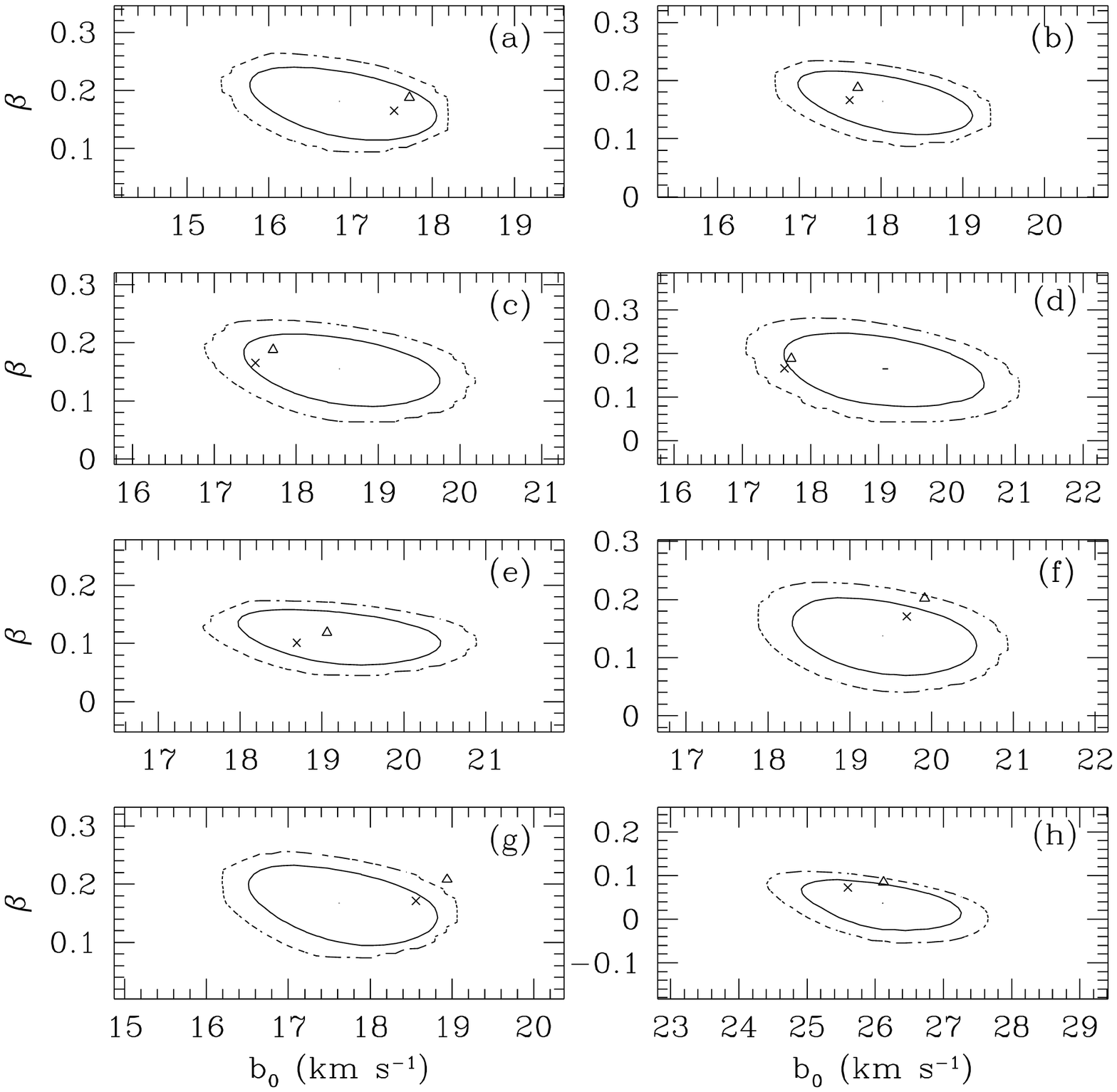}
\caption{\label{fig:6m}\capsm}
\end{figure}
}
In Fig.~\ref{fig:6m} we present the result with which we are more
concerned in this paper: that the parameters of the equation of state
can be measured from the $b-N$ distribution within the precision of
the method. Each panel shows the best fit (black dot) and the
$1\sigma$ and $2\sigma$ error contours for $b_0$ and $\beta$. The
triangle and the cross are the theoretical values derived using \eq
(\ref{eq:bN}) for two slightly different measurements of the
parameters $N_{\overline \rho}$ and $\Gamma$ in the \eq (\ref{eq:Nd}).
In one case, we fit the $N-\rho$ distribution with constant weighting
for all the lines (triangles), and in the other (crosses) with a
$N^{1/2}$ weighting in order to correct for the effect of the
decreasing number of lines with increasing the column density.  For a
more detailed explanation see \S~\ref{ssec:n-d}. In all the models,
the expected values of $b_0$ and $\beta$ lie near the $1\sigma$
contour level of the measurement, as they should be if the errors are
estimated correctly.

%6d-is the decrease in b real or not? .....$N_{\overline \rho}$ is close to the maximum
%       of the distribution but there are several uncertain like
%       blending and resolution.
%6e-different redshift result?
%find phi=phi1+phi2*N

\subsection {Relation Between Column Density and Overdensity\label{ssec:n-d}}

The Density Peak Ansatz (DPA) approximation is based on the assumption
 that a single \lya line arises from a peak in the cosmic gas
 density along the line of sight to a distant quasar. Then, given the
 value of the cosmic density at the peak, and the
 second derivative of the density along the line of sight (the first
 derivative vanishes at the peak), a value for the column density can
 be associated with the peak as follows (\markcite{gh96}Gnedin \& Hui 1996):
\[
        N_{\HI} \equiv (3.35\times10^{13}\dim{cm}^{-2})T_{0,4}^{-0.7}
        \left(\Omega_bh^2\over0.019\right)^2\left(0.5\over J_{21}
        \right)\left(1+z\over4\right)^5
\]
\begin{equation}
        \ \     
        \times (1+\delta)^{2-0.7(\gamma-1)}
        \left[-(1-0.35(\gamma-1)){d^2\dim{ln}(1+\delta)\over dx^2}
        \right]^{-1/2},
        \label{eq:DPA}
\end{equation}
and $x$ being measured in Mpc. Here $J_{21}$, is the ionizing intensity,
\[            
        J_{21} \equiv
        {\int J_\nu \sigma_\nu d\nu/\nu\over \int \sigma_\nu d\nu/\nu}
        \times
        {1\over10^{-21}\dim{erg}\dim{cm}^{-2}\dim{s}^{-1}\dim{Hz}^{-1}
        \dim{sr}^{-1}},
        \label{J21def}
\]            
where $J_\nu$ is the radiation specific intensity and $\sigma_\nu$ is
the \HI\ photoionization cross section.
%\def \capre{% 
%Typical comoving size of a \lya cloud (see \eq
%[\ref{eq:xi}]), as a function of the linear density fluctuation at the
%filtering scale $\sigma_F$ for 25 different cosmological models. (a)
%Triangles show $\xi$ versus $\sigma_F$ for 25 cosmological models
%(Gnedin 1998)  and the solid line corresponds to the best fit. (b)
%Triangles show $\Xi$ versus $\sigma_F$ for 25 models and the
%solid line corresponds to the best fit. Parameters of the best fit are
%shown in Table \ref{tab:xi}.} 
%\placefig{
%\begin{figure}
%\epsscale{0.50}
%\inserttwofigures{relation1.eps}{relation2.eps}
%\caption{\label{fig:re}\capre}
%\end{figure}
%}
Comparing \eqs (\ref{eq:Nd}) and (\ref{eq:DPA}) we find that one can
write the inverse of the square root of the average curvature at a
density peak, \ie\ the typical comoving size (expressed in Mpc) of a
\lya cloud, as a function of the overdensity,
\begin{equation}
\left(-{d^2\dim{ln}(1+\delta)\over dx^2}\right)^{-1/2}=\Xi(1+\delta)^\xi,
\label{eq:xi}
\end{equation}
where $\Xi$ and $\xi$ are model dependent. We hypothesize that the
average curvature of a density peak is a function of the non-linearity
of the baryonic density field.  An accurate indicator of the
non-linearity of the gas distribution of a given cosmological model is
the linear density fluctuation at the filtering scale, $\sigma_{F}$
(\markcite{gh98}Gnedin \& Hui 1998). 
%In Fig.~\ref{fig:re} we show
%$\ln \Xi$ and $\xi$ versus $\sigma_{F}$ for 25 flat cosmological
%models (\markcite{gne98}Gnedin 1998). The solid lines are the best
%linear fit to the points, and the result of the fit is given in
%Table~\ref{tab:xi}.
We therefore fitted $\Xi$ and $\xi$ as linear function of $\sigma_F$
for 25 flat cosmological models (\markcite{gne98}Gnedin 1998). 
The result of the fit is given in Table~\ref{tab:xi}.
The parameters in \eq (\ref{eq:Nd}) are measured
using the minimum of Chi-square analysis of the $N-\rho$ distribution
with a weighting on each line $\propto N^{1/2}$ in order to account
for the $0.5$ slope of the logarithm of the column density
distribution.
%
% Table 5
%
\def\tabthree{
\begin{deluxetable}{lccc}
\tablecaption{Parameters of the best fit \label{tab:xi}}
\tablehead{
\colhead{}   & \colhead{$a$} & \colhead{$b$} & \colhead{$\chi^2$}
 }
 \startdata
 $\Xi=a+b \sigma_F$ & $0.19 \pm 0.02$ & $-0.025 \pm 0.01$ & 0.005  \nl
 $\xi=a+b \sigma_F$ & $-0.319 \pm 0.02$ & $0.068 \pm 0.01$ & 0.013  \nl
 \enddata
\end{deluxetable}}
\placefig{\tabthree}

Combining \eqs (\ref{eq:xi}) and (\ref{eq:DPA}), we write the
relationships,
\[
\Gamma=2-0.7(\gamma-1)+\xi \simeq
1.68 + 0.068 \sigma_F - 0.7(\gamma-1),
\]
\begin{equation}
N_{\overline \rho}=(3.35\times10^{13}\dim{cm}^{-2})T_{0,4}^{-0.7}
        \left(\Omega_bh^2\over0.019\right)^2\left(0.5\over J_{21}
        \right)\left(1+z\over4\right)^5 {0.19-0.025\sigma_F \over [1-0.35(\gamma-1)]^{1/2}}.
\label{eq:Nrho}
\end{equation}
Both $\Gamma$ and $N_{\overline \rho}$ are insensitive to the value of
$\sigma_{F}$ if we make a conservative assumption that it is in the
range $2 \pm 1$ (\markcite{gne98}Gnedin 1998). From now on, we use a
value $\sigma_{F}=2$.

Finally, if we define $N_{\overline \rho}^*=N_{\overline
\rho}T_4^{0.7}$, we can write the equations that relate the measured
parameters $\beta$ and $T_{0,4}^\dag$ to the \eos\ parameters,
\begin{eqnarray}
\gamma-1 &\simeq& {3.634 \beta \over 1+1.4 \beta},\label{eq:gbeta} \\
T_{0,4}&=&\left[T_{0,4}^\dag \left({N_{\overline \rho}^* \over N_0}\right)^{2
\beta}\right]^{1 \over 1+1.4 \beta},
\label{eq:bt4}
\end{eqnarray}
where,
\begin{equation}
{N_{\overline \rho}^* \over N_0} \simeq
        {(4.7\times10^{12}\dim{cm}^{-2}) \over N_0}
        \left(\Omega_bh^2\over0.019\right)^2\left(0.5\over J_{21}
        \right)\left(1+z\over4\right)^5 \sqrt{1+1.4 \beta \over
        1+0.128 \beta}.
\end{equation}
Propagating the errors on $\beta$ and $b_0$, we write the errors on the
\eos\ parameters,
\begin{eqnarray}
{\Delta (\gamma-1) \over \gamma-1} &=& {1 \over 1+1.4 \beta}{\Delta
\beta \over \beta},    \\
{\Delta T_{0,4} \over T_{0,4}} &\simeq& {2 \over 1+1.4 \beta}\left(\left|{\Delta b_0 \over
b_0}\right|^2+\left|\left(\ln {N_ {\overline \rho}^* \over N_0}-1.4\ln
{b_0 \over 13}\right){\Delta
\beta \over 1+1.4\beta} \right|^2\right)^{1 \over 2}.
\label{eq:err}
\end{eqnarray}
If $N_{\overline \rho}^*/N_0=(b_0/13~\rm{km s}^{-1})^{1.4}$, the errors are
uncorrelated and $\Delta T_{0,4}$ has a minimum value.

\section{Observed Evolution of the Equation of State \label{sec:z-evol}}

\subsection {Observations \label{ssec:obs}}

We apply our method to the published data sets of \lya absorption
lines. In this paper, we analyze eight data sets, but the \eos\ can be
measured with much better accuracy if we use more observations. In
order to obtain good results, we need to use the data sets with small
errors on $b$ and $N$ and with as many lines as possible inside the range
of the column densities where the completeness of the sample is good.  For
the high-redshift \lya forest we use Keck-HIRES observations of the
quasars Q0000-26 (\markcite{lu96}Lu \etal\ 1996) and APM
08279+5255\footnote{The data presented herein were obtained at the
W. M. Keck Observatory, which is operated as a scientific partnership
among the California Institute of Technology, the University of
California and the National Aeronautics and Space Administration.  The
Observatory was made possible by the generous financial support of the
W. M. Keck Foundation.} (\markcite{ell99}Ellison \etal\ 1999) at $z
\sim 4$, Q0014+813, Q0302-003, Q0636+680, Q0956+122 (\markcite{hu95}Hu
\etal\ 1995), GB1759+7539 (\markcite{out99}Outram \etal\ 1999) and HS
1946+7658 (\markcite{kir97}Kirkman \& Tytler 1997) at $z \sim 3$. At
intermediate redshift we use the data-set of \markcite{kha97}Khare
\etal\ 1997 for the quasar B2 1225+317 at $z \sim 1.9$. Finally, we
use HST observations of the local \lya forest kindly provided by
\markcite{pen99}Penton, Stocke, \& Shull (1999), using the Goddard
High Resolution Spectrometer (GHRS) with 19 km s$^{-1}$ resolution.

%
% Table 4
%
\def\tabfour{
\begin{deluxetable}{lccccc}
\tablecaption{Observational data \label{tab:o1}}
\tablehead{
\colhead{QSO}   & \colhead{Telescope} & \colhead{FWHM} &
\colhead{S/N} & \colhead{Voigt fitting} &
\colhead{number}\\[0.5ex]
\colhead{} & \colhead{} & \colhead{km s$^{-1}$} & \colhead{} & \colhead{code} &
\colhead{of lines}
 }
 \startdata
 Q0000-26 & Keck HIRES & 6.6 & 20-80 & VPFIT & 337 \nl
 APM 08279+5255 & Keck HIRES & 6 & 80 & ... & 380 \nl
 Hu \etal $^a$ & Keck HIRES & 8 & 50 & ... & 1056 \nl
 GB 1759+7539  & Keck HIRES & 7 & 25 & ... & 318 \nl
 HS 1946+7658 & Keck HIRES & 7.9 & 15-100 & VPFIT & 459 \nl
 B2 1225+317 & Kitt Peak 4-m & 18 & 2-20 & ... & 159 \nl
 Penton \etal $^b$  & HST/GHRS & 19 & 10-42 & ... & 80 \nl
 \enddata
 \tablecomments{The data for the quasars at high-$z$ are from Lu
\etal\ 1996 (Q0000-26), Ellison \etal\ 1999 (APM 08279+5255), Outram
\etal\ 1999 (GB1759+7539), Kirkman \& Tytler 1997 (HS 1946+7658), Khare \etal\ 1997 (B2 1225+317).}
  \tablenotetext{a}{Collection of high-$z$ quasars Q0014+813, Q0302-003,
  Q0636+680, Q0956+122 (Hu \etal, 1995).}
  \tablenotetext{b}{Collection of low-$z$ quasars: AKN120,
  ESO141-G55, FAIRALL9, MARK279, MARK290, MARK421, MARK509, MARK817,
  MARK501, PKS2155, Q1230+0115, H1821+643, \hbox{ESO141-G55} (Penton \etal\ 1999).}
\end{deluxetable}
}
\placefig{\tabfour}
\def\caplu{% 
(a) Distribution of $\log b$ versus $\log N$ for \lya lines for
Q0000-26 at $\langle z \rangle \sim 3.7$, with errorbars from
VPFIT. (b) Comparison between the $b$ distribution for Q0000-26
(long-dashed line) and for model LCDM.3A (solid line) at $z=2.8$ and
$\overline N \simeq 10^{13}$ \cm2. The large errors on $b$ and $N$ of
the observed \lya forest spread considerably the true distribution.}
\placefig{
\begin{figure}
%\epsscale{0.50}
\inserttwofigures{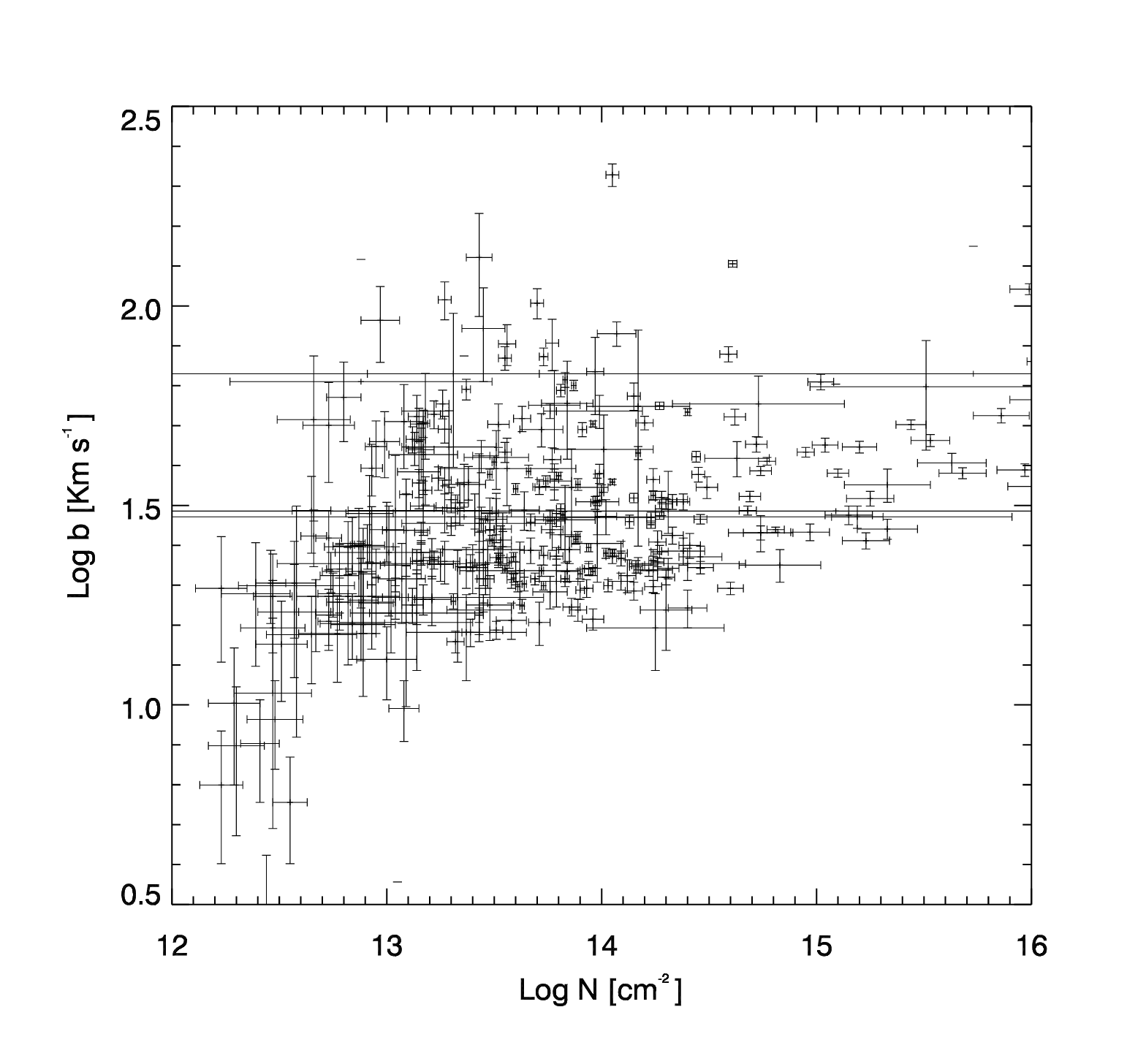}{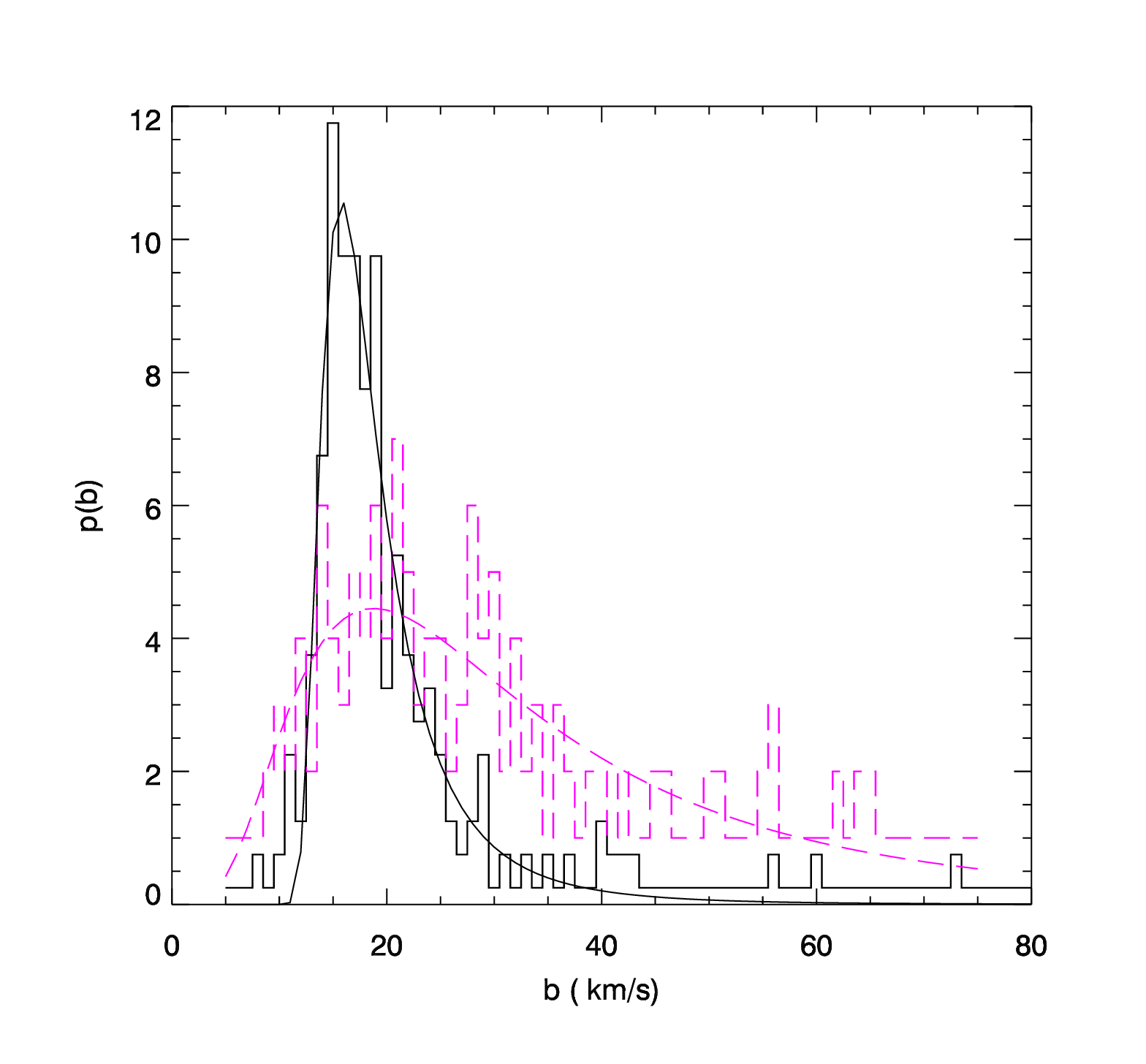}
\caption{\label{fig:lu}\caplu}
\end{figure}
}
We use a subsample of 337 \lya absorption lines of the complete Lu
\etal\ (1996) data set within a redshift range $3.425 \le z \le
3.977$. The upper cutoff is chosen because this is where the UV
ionizing intensity from the quasar is expected to roughly equal that
of the UV background.
  
The data set from Hu \etal\ (1995) is a collection of four quasars and
consists of around $10^3$ lines in the redshift range $3.425 \le
z \le 3.977$.  Unfortunately, the published list of lines does not have
errors on the $b$ parameter and the column density. In order to use this
large list of lines, we decided to produce synthetic errors, assuming that
their amplitude and distribution follows the same statistics as the Lu
\etal\ (1996) sample. We chose this sample because it has similar
resolution and S/N. Unfortunately, the Voigt profile fitting routine is
not the same.

The local \lya data set (Penton \etal\ 1999) consists of 79 lines from
15 Seyfert galaxies and quasars detected in observations with the
HST/GHRS, using both pre- and post-COSTAR
optics. We do not use the \hbox{pre-COSTAR} observations because this
sample appears to have systematically larger $b$-values due to
spacecraft wobble (Penton \etal\ 1999). This reduces the sample to 43
lines, which are distinct \lya absorption features at $>4 \sigma$
significance, with $\langle b \rangle=35.0 \pm 16.6$ km s$^{-1}$.
At low redshift, a typical line of sight to a QSO at
$z\sim0.05$ may have fewer than 10 absorbers. As a result, it is not
possible to have an uniform sample of lines belonging to the same line
of sight such as for the high-redshift \lya forest.
However, the ensemble of low-$z$ \lya absorbers, drawn from 15
sight-lines, adequately characterizes the distribution of $b$-values down
to the 19 km s$^{-1}$ spectral resolution of the HST/GHRS. Additional
\lya lines are currently being observed with the Space Telescope
Imaging Spectrograph (STIS) and will double the sample of lines.

In all the samples, we do not include identified metal lines.  We only
use absorption lines in a limited range of the column densities. The lower
limit is determined by imposing the following two conditions:
\begin{itemize}
\item the completeness of the sample must be close to one;
\item the width of the $b$ distribution must be monotonically
decreasing in the range considered.
\end{itemize}
The first condition is usually
the most stringent. In the case of Q0000-26, the two conditions are
comparable. The upper limit on the column density is set by imposing
the following constraints:
\begin{itemize}
\item overdensity of the line $\delta \approxlt 10$ (range of validity
of the HPM approximation).
\item $N_{max} \le 10^{14.5}$ \cm2\ in order to avoid heavily saturated
lines.
\end{itemize}
By restricting ourselves to this range of the column densities, we are
more confident of the reliability of the results from the Voigt
profile fitting algorithm. Indeed, we find that the shapes of the
$b$-distributions obtained by different authors, using different Voigt
profile fitting algorithms, are very similar in this range of the
column densities. It is however possible that fluctuations of the
ionizing background (\markcite{haa96}Haardt \& Madau 1996; Fardal
\etal\ 1998) can produce different mean absorption of the quasar
spectra and result in a spatially fluctuating value of $N_{\overline
\rho}$. However, this effect is much smaller than the errors of the
measurement. We will quantify the change of the IGM temperature
produced by a change in the ionizing background below.

We discard lines with errors on $b$ or $N$ greater than $50\%$ in
order to obtain more precise results. Large errors degrade the
precision of the final result. On the other hand, the number of lines
of the sample decreases if we consider only the lines with small
errors. By performing several MLAs, with various error thresholds, we
find that limiting the relative errors to about $50\%$ is the best
choice.

In Table~\ref{tab:o1} we summarize the main properties of the data
sets used in this paper. To give an idea of the magnitude of the errors
and how they can affect the shape of the $b$ distribution, we show in
Figs.~\ref{fig:lu}a and \ref{fig:lu}b the data set of Lu
\etal\ (1996), with error-bars and the $b$ distribution at $\overline N\sim
10^{13}$ \cm2\ compared to a typical simulated $b$ distribution.

\subsection {Results\label{ssec:results}}
%
% Table 4
%
\def\tabfive{
\begin{deluxetable}{lcccccccccc}
\footnotesize
%\tablewidth{\textwidth}
\tablecaption{Observed Ly-$\alpha$ forest \label{tab:o2}}
\tablehead{
\colhead{QSO}   & \colhead{$\langle z \rangle$} & \colhead{number} & \colhead{log N$_{min}$} & \colhead{log N$_0$} &
\colhead{b$_0$} & \colhead{$\beta$} & \colhead{$\varphi_0$}
 & \colhead{$\varphi_1$} & \colhead{$\Phi$} & \colhead{$\alpha$}\\[0.5ex]
\colhead{} & \colhead{} & \colhead{of lines$^c$} & \colhead{(cm$^{-2}$)}  & \colhead{(cm$^{-2}$)} &
\colhead{(km s$^{-1}$)} & \colhead{}    & \colhead{} &
& \colhead{} & \colhead{}
 }
 \startdata
 (a) Q0000-26 & 3.7 & 195 & 13.5 & 14.15 & 23.6 & 0.11 & 3.6 & 0.1 &
4.6 & 1.7 \nl
 (b) APM 08279+5255 & 3.42 & 186 & 13.0 & 13.7 & 23.1 & 0.09 & 2.6 &
0.3 & 5.6 & 1.7 \nl
 (c) Hu \etal $^a$ & 2.85 & 636 & 13.0 & 13.7 & 24.3 & 0.04 & 4.6 & 0.01 & 3.1 & 1.6\nl
 (d) Q0014+813 & 2.95 & 159 & 13.0 & 13.7 & 26.4 & 0.07 & 5.9 & 0.1 & 2.1 & 1.7 \nl
 (e) Q0302-003 & 2.85 & 160 & 13.0 & 13.7 & 26.9 & 0.11 & 4.8 & 0.2 & 2.8 & 1.7 \nl
 (f) Q0636+680 & 2.75 & 193 & 12.8 & 13.7 & 23.9 & 0.05 & 4.7 & 0.06 & 3.3 & 1.6 \nl
 (g) Q0956+122 & 2.85 & 180 & 12.8 & 13.7 & 25.7 & 0.04 & 4.3 & 0.02 &
4.4 & 1.6 \nl
 (h) GB 1759+7539 & 2.72 & 13.5 & 13.9 & 13.7 & 23.2 & 0.16 & 3.1 & 1.5 & 2.0 & 2.1 \nl
 (i) HS 1946+7658 & 2.7 & 221 & 13.0 & 13.6 & 22.4 & 0.06 & 3.3 & 0.05 & 7.2 & 1.8 \nl
 (l) B2 1225+317 & 1.9 & 99 & 13.2 & 13.7 & 24.4 & 0.1 & 6.2 & 0.5 & 1.2 & 1.8 \nl
 (m) Penton \etal $^b$ & 0.06 & 43 & 12.5 & 13.48 & 34.2 & 0.35 & 12.4 & 0.7 & 1.4 & 1.6 \nl
 \enddata
\tablenotetext{a, b}{ See notes on Table~\ref{tab:o1}.}
\tablenotetext{c}{ Number of lines used in the MLA in the
column density range $N_{min} \le N \le N_{max}$.}
\end{deluxetable}
}
\placefig{\tabfive} 
In Table~\ref{tab:o2} we report the parameters
for the best fit from the MLA. We use the same form of the function as
adopted in the simulations (see \eq [\ref{eq:pnb}]).
Fig.~\ref{fig:o1} shows the values of $b_0$ and $\beta$ and the
error contours at $1\sigma$ and $2\sigma$ confidence levels. Each
panel shows the results for a different quasar or data set. In
Table~\ref{tab:o2} the quasar and the mean redshift of the \lya forest
are reported along with the label that appears on the upper right
corner of each panel.  Fig.~\ref{fig:o2} is analogous to
Fig.~\ref{fig:o1} but shows the value and error contours for
$\gamma-1$ and $T^\dag_{0,4}$ derived using \eq
(\ref{eq:gbeta}). Finally in Fig.~\ref{fig:9r} we show the values and
the $1\sigma$ error contour for $\gamma-1$ and $T_0$ (computed with
\eq [\ref{eq:bt4}]). In each panel we have the result for two
different values of the ionizing intensity $J_{21}$. The values chosen
for $J_{21}$ at different redshifts are based on observational (Lu
\etal\ 1996; \markcite{shu99}Shull \etal\ 1999) and theoretical
(\markcite{haa96}Haardt \& Madau 1996; \markcite{far98}Fardal \etal\
1998; \markcite{val99}Valageas \& Silk 1999) results. We adopt as
minimum values of $J_{21}$:
$J_{21}^{min}(z=4)=0.2,~J_{21}^{min}(z=3)=0.5,~J_{21}^{min}(z=2)=0.2$
and $J_{21}^{min}(z=0)=0.01$. The maximum values are,
$J_{21}^{max}(z) =2 \times J_{21}^{min}(z)$ for all $z$.

\def\capoa{% 
Error ellipses in the $\beta \otimes b_0$ space for the
data sets listed in Table \ref{tab:o2}. Solid lines show
$1\sigma$ error contours and the dashed lines show the $2\sigma$ error
contours.}
\placefig{
\begin{figure}
%\epsscale{0.50}
\insertfigure{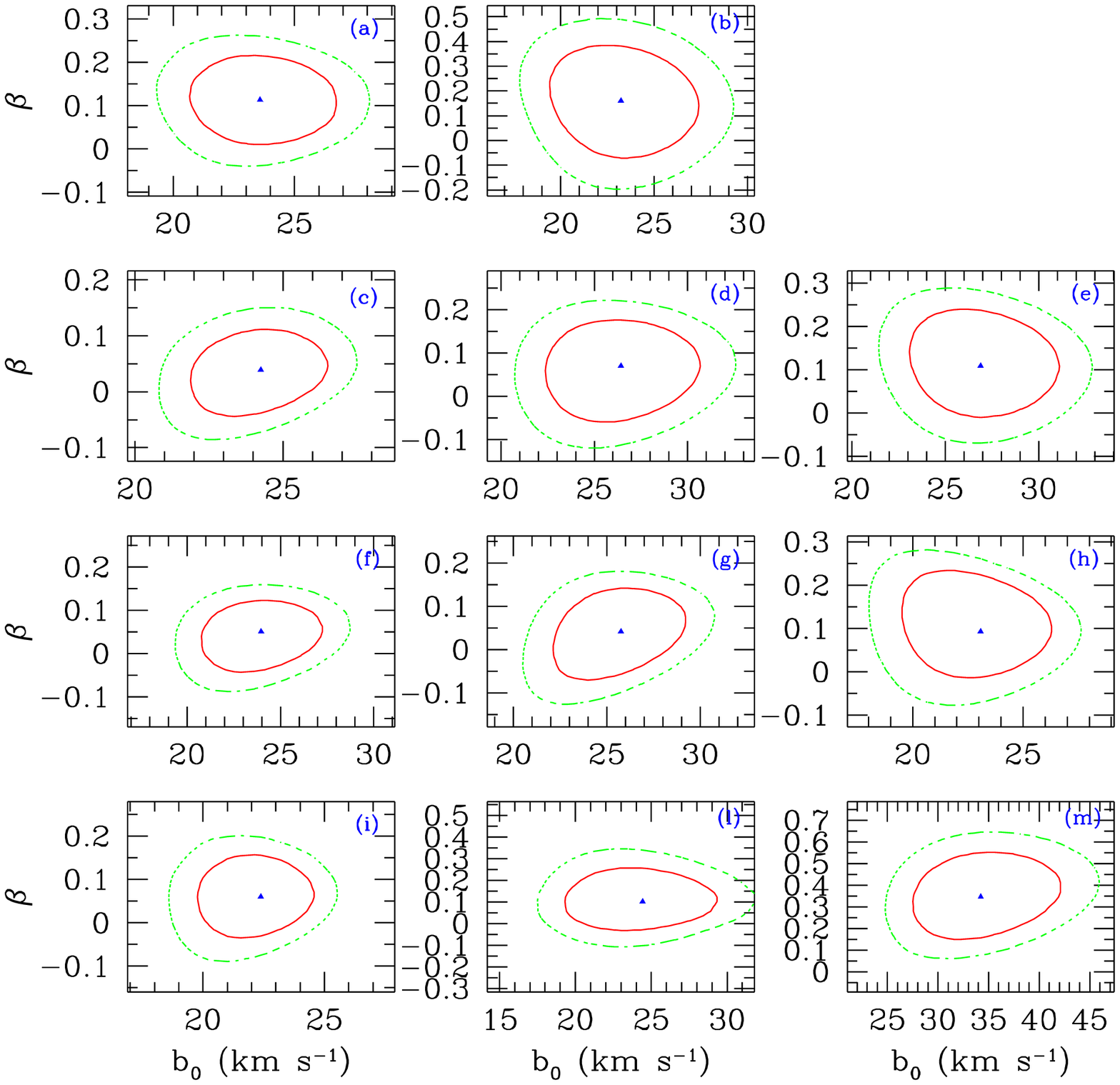}
\caption{\label{fig:o1}\capoa}
\end{figure}
}
\def\capob{%
Error ellipses in the $\gamma-1 \otimes \log T_0^\dag$ space for the
data sets listed in Table \ref{tab:o2}. Solid lines show
$1\sigma$ error contours and the dashed lines show the $2\sigma$ error
contours.}
\placefig{
\begin{figure}
%\epsscale{0.50}
\insertfigure{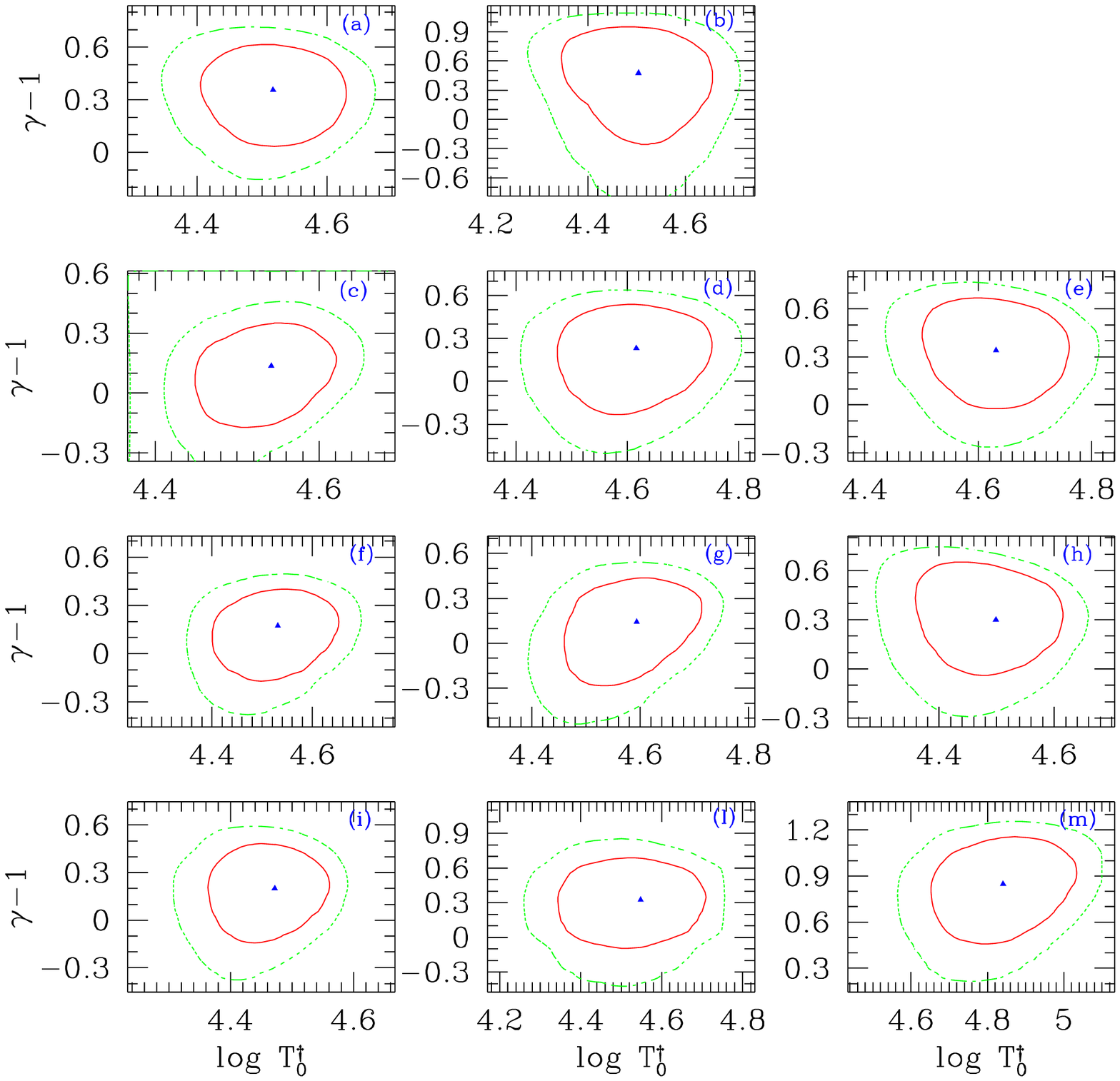}
\caption{\label{fig:o2}\capob}
\end{figure}
}
\def\cap9r{%
$1\sigma$ error ellipses in the $\gamma-1 \otimes \log T_0$ space for the
data sets listed in Table \ref{tab:o2}. The two ellipses in each
panel are calculated using $J_{21}=J_{21}^{min}, J_{21}^{max}$. Values
are reported in the text.}
\placefig{
\begin{figure}
%\epsscale{0.50}
\insertfigure{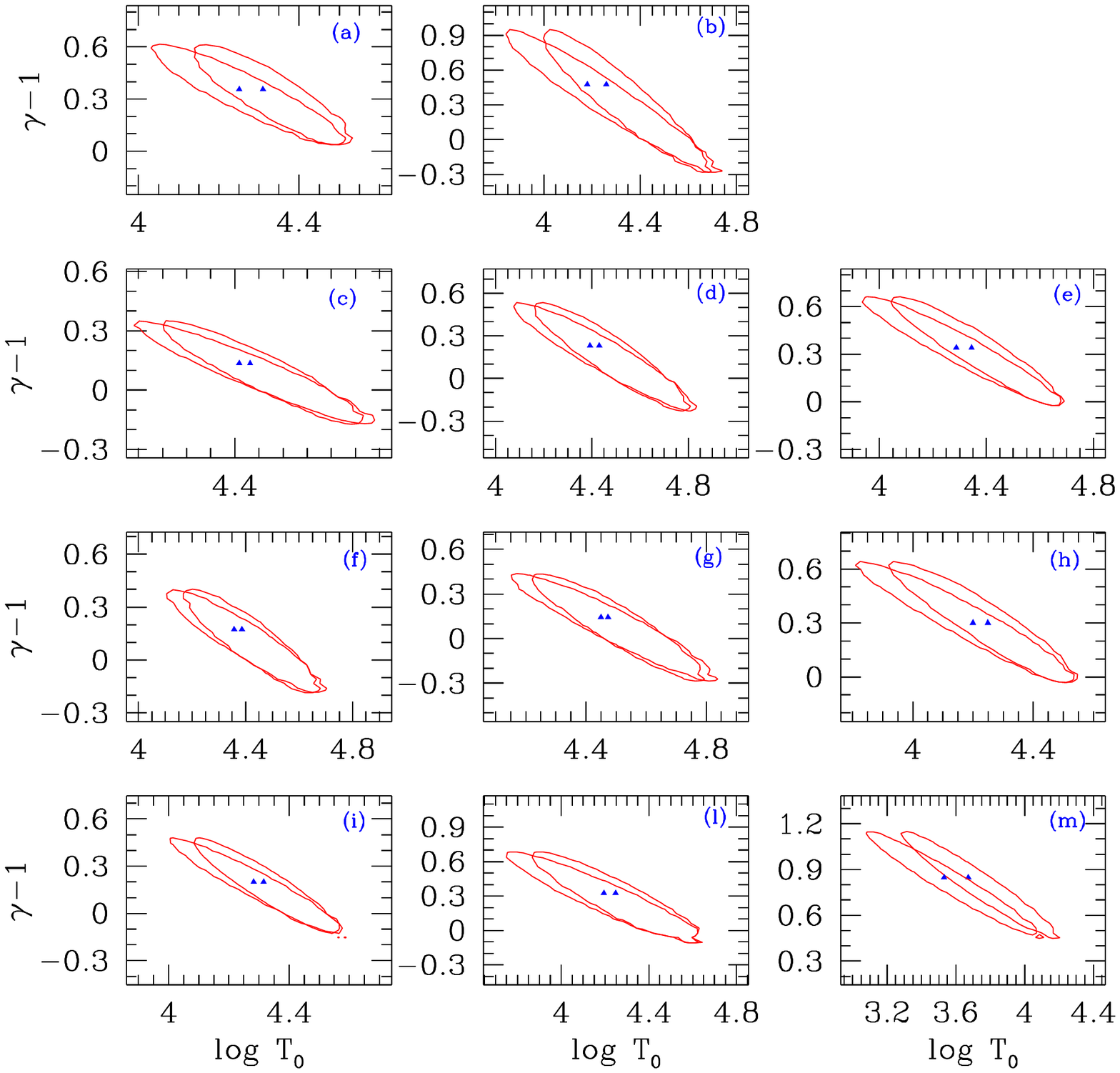}
\caption{\label{fig:9r}\cap9r}
\end{figure}
}
The error ellipses are strongly elongated and tilted with respect to
the Cartesian axis. This happens because $N_{\overline \rho}^* \not=
N_0$.  In this case, as shown in \eq (\ref{eq:err}), the error on
$T_0$ depends on the value of $\gamma-1$. It is impossible to remove
this correlation of the errors by changing the value of the
normalization parameter $N_0$ because this would introduce exactly the
same amount of correlation from the errors on $b_0$ and $\beta$ due to
the limited range of the column densities of the lines.  Using a
smaller value of the minimum column density of the distribution of
absorption lines would be the only way to reduce this
correlation. This means that, in order to have $N_{\overline
\rho}^*/N_0=1$, we would need a data set that is also complete for
lines arising from the density fluctuations in the underdense regions
($\delta < 0$). This would probably imply the use of a more
complicated form of the parametric function if the $b$ distribution
starts to become narrower in underdense regions as can be expected
on theoretical grounds.

%The values of $N_{\overline \rho}$ obtained from \eq \ref{eq:Nrho}
%are $N_{\overline \rho}=6\times 10^{11}, 1.6\times 10^{12},
%2.3\times10^{12}, 1.5 \times 10^{13}$ \cm2 at $z \sim 0, 2, 3, 4$
%respectively. This values are determined by the assumed ionizing
%background $J_{21}$, baryonic fraction $\Omega_b$, Hubble constant and
%are slightly dependent on the power spectrum of the density field
%(related to $\sigma_F$) as has been also noted by \markcite{br99}Bryan
%\& Machacek (1999). We expect that at the column density $N \approxlt
%N_{\overline \rho}$ the shape of the $b$-distribution changes,
%\ie\ the power law tail should start to narrow. This change in the
%shape of the distribution is observed to occur at $N \sim 1-5 \times
%10^{13}$ \cm2 at redshift $z \sim 4$ and at $N \approxlt 5 \times 10^{12}$
%\cm2 at $z < 3$. The value of $N_{\overline \rho}$ we use are consistent
%with this observed behavior. \markcite{da99}Dav\'e \etal\ (1999), found
%values of $N_{\overline \rho}=1.6\times 10^{12}, 4\times 10^{12},
%1.6\times10^{13}, 10^{14}$ \cm2 at $z=0, 1, 2, 3$ respectively. This
%values are bigger because they use a higher baryonic fraction and the
%size $\Xi$ of the clouds is bigger. Adopting their values of
%$N_{\overline \rho}$ would imply a decrease of the IGM
%temperature. Anyway this would not affect very much the temperature at
%$z \sim 3$ where the gas is almost isothermal.

\def\capfi{%
The evolution of $\gamma$ ({\it a\/}) and $T_0$ ({\it b\/}) with
redshift. The point
with error bars at $z \sim 4$ is the weighted average of the
observations (a) and (b), and the point at $z \sim 3$ is the weighted
average of the observations (d), (e), (f), (g), (h), (i), where the
letters refer to the list in Table~\ref{tab:o2}. The solid line
represents the theoretical model with a sudden \HI\ reionization at
$z_{rei}=7$ and with $T_{\rm rei}=2.5 \times 10^4$ K. The dashed lines
show the effect of sudden \GII\ reionization in the optically thin
approximation ({\it short-dashed line\/}) and with the \GII\
photoheating rate quadrupled ({\it long-dashed line\/}). The model
shown by the dot-dashed line has a quadrupled \GII\ heating rate and a
more gradual \GII\ reionization.}
%The
%value of $\gamma-1$ at $z\sim 3$ is very low and the temperature at $z
%\sim 3$ is too high, only marginally consistent with the sudden \HI
%reionization model. We interpret this discrepancy as a secondary
%reheating due to the reionization of \GII\ observed to occur at $z\sim3$.}
\placefig{
\begin{figure}
%\epsscale{0.60}
\inserttwofigures{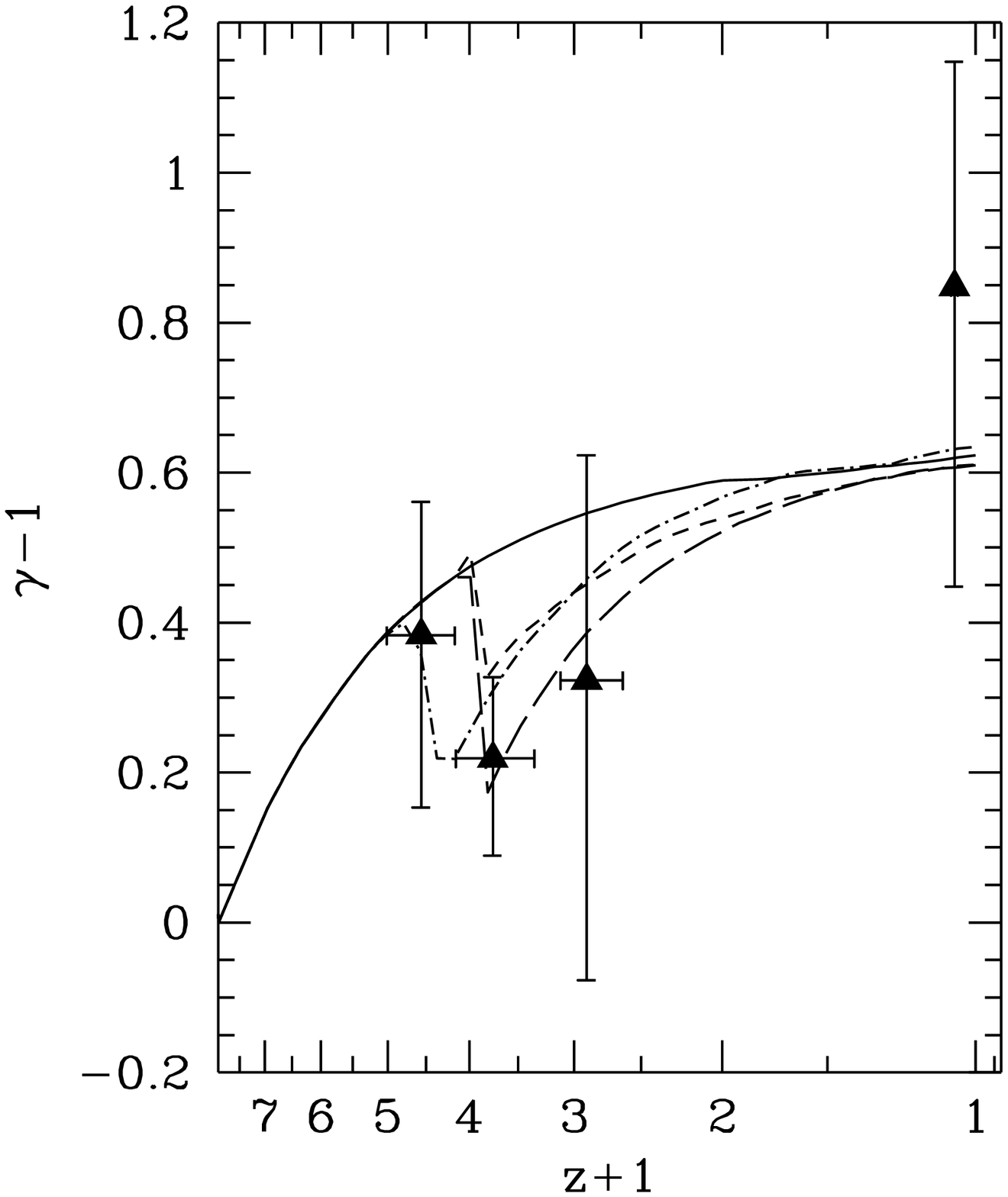}{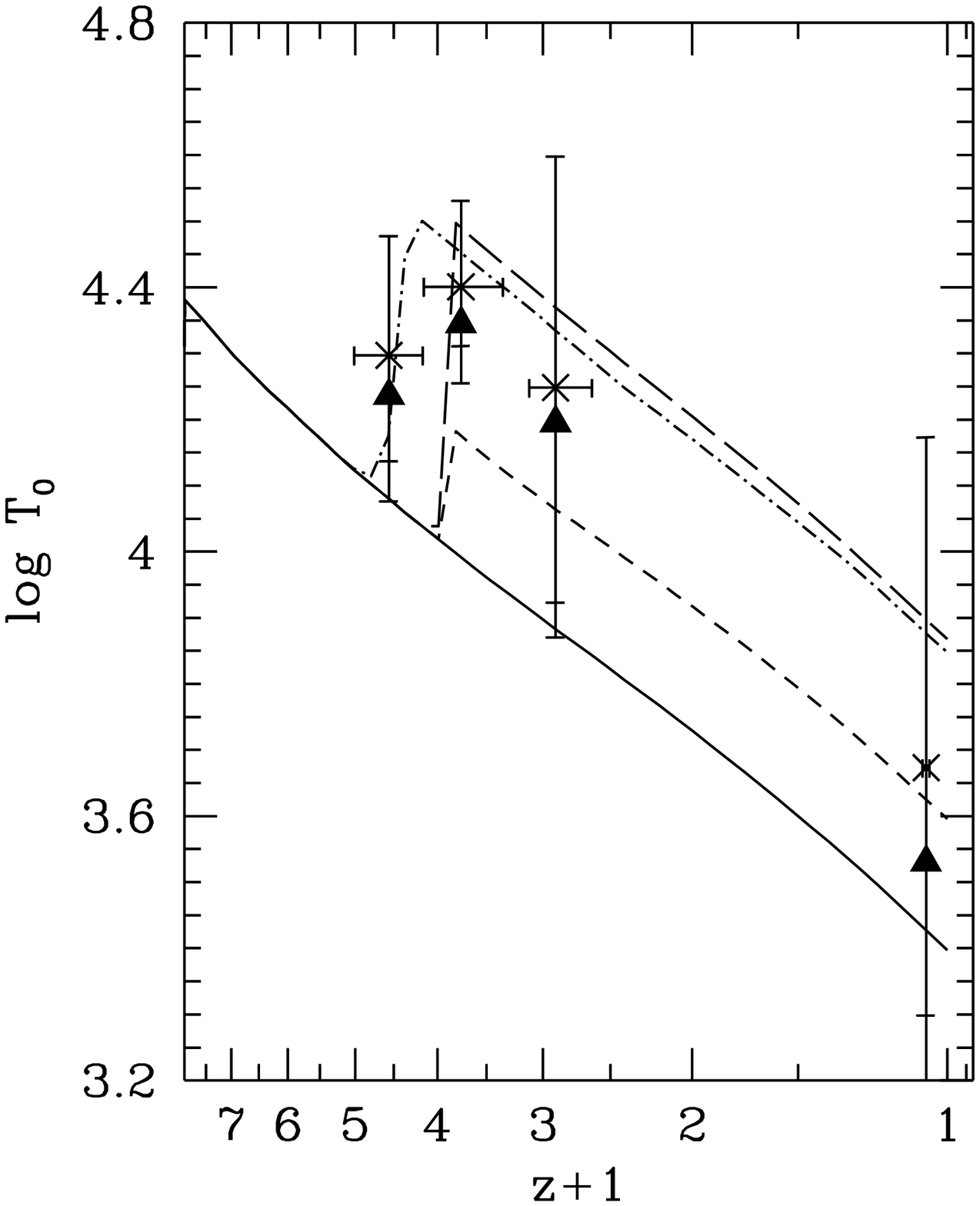}
\caption{\label{fig:fi}\capfi}
\end{figure}
} 
In Fig.~\ref{fig:fi}, we show the final result of this work, the
evolution of the \eos\ of the IGM with $1\sigma$ error-bars and for
two different values of the ionizing radiation $J_{21}$, chosen as
explained above. The results are also listed in Table~\ref{tab:f1} and
Table~\ref{tab:f2} for
reference.  The point at $z \sim 4$ is the weighted average of the
observations (a) and (b), and the point at $z \sim 3$ is the weighted
average of the observations (d), (e), (f), (g), (h), (i), where the
letters refer to the list in Table~\ref{tab:o2}.

The lines in Fig.~\ref{fig:fi} show simple photoionization models 
for the thermal history of the universe computed using the method of
Hui \& Gnedin (1998). In all the models, hydrogen is reionized suddenly
at $z=7$. The solid line shows the model where \GII\ is not reionized at all,
the short-dashed lines show models with sudden helium reionization
at $z=3$, the long-dashed line shows the same model but with
the \GII\ photoheating rate artificially increased by a
factor of four following the suggestion by
Abel \& Haehnelt (1999). Finally, the dot-dashed line shows a model where
the \GII\ reionization is more gradual, and it provides a remarkably good
fit to the data.

%
% Table 6
%
\def\tabsix{
\begin{deluxetable}{lccccc}
\tablecaption{Time evolution of the observed \eos. \label{tab:f1}}
\tablehead{
\colhead{$\langle z \rangle$} & \colhead{$\gamma-1$} & \colhead{$T_0/10^4$ K}
& \colhead{$T_0/10^4$ K ($J_{21} \times 2$)} & \colhead{$\Delta T_0/T_0$} & \colhead{$J_{21}$}
 }
 \startdata
 $3.56 \pm 0.27$ & $0.38\pm 0.18$ & 1.98 & 1.72 & $16\%$ & 0.2 \nl
 $2.75 \pm 0.27$ & $0.22\pm 0.10$ & 2.52 & 2.21 & $9\%$  & 0.5 \nl
 $1.90 \pm 0.30$ & $0.32\pm 0.30$ & 1.77 & 1.57 & $32\%$ & 0.2 \nl
 $0.06 \pm 0.01$ & $0.85\pm 0.30$ & 0.47 & 0.34 & $37\%$ & 0.01\nl
 \enddata
\end{deluxetable}
}
\placefig{\tabsix}
%
% Table 7
%
\def\tabseven{
\begin{deluxetable}{lccccc}
\tablecaption{Measured \eos\ of the observations at $z \sim 4$ and $z \sim 3$. \label{tab:f2}}
\tablehead{
\colhead{$\langle z \rangle$} & \colhead{$\gamma-1$} & \colhead{$T_0/10^4$ K}
& \colhead{$T_0/10^4$ K ($J_{21} \times 2$)} & \colhead{$\Delta T_0/T_0$} & \colhead{$J_{21}$}
 }
 \startdata
 $3.70 \pm 0.27$ (a) & $0.35\pm 0.2$ & 2.04 & 1.78 & $15\%$ & 0.2 \nl
 $3.42 \pm 0.27$ (b) & $0.47\pm 0.4$ & 1.78 & 1.51 & $25\%$  & 0.2 \nl
 $2.95 \pm 0.25$ (d) & $0.23\pm 0.3$ & 2.69 & 2.47 & $30\%$ & 0.5 \nl
 $2.85 \pm 0.25$ (e) & $0.34\pm 0.3$ & 2.21 & 1.94 & $30\%$ & 0.5 \nl
 $2.75 \pm 0.25$ (f) & $0.17\pm 0.2$ & 2.43 & 2.27 & $20\%$ & 0.5 \nl
 $2.85 \pm 0.25$ (g) & $0.14\pm 0.3$ & 2.97 & 2.82 & $25\%$ & 0.5 \nl
 $2.72 \pm 0.33$ (h) & $0.30\pm 0.3$ & 1.77 & 1.58 & $35\%$ & 0.5 \nl
 $2.70 \pm 0.30$ (i) & $0.14\pm 0.25$ & 2.69 & 2.56 & $20\%$ & 0.5 \nl
 \enddata
\end{deluxetable}
}
\placefig{\tabseven}

\subsection{Sanity Check}

We would like to emphasize that the thermal history of the universe
which we derive from the observational data is quite different from
the one assumed in all the simulations we used to test the method: the
temperature is significantly higher and the \eos\ flattens out at
$z=3$.  As a sanity check, and in order to test the reliability of the
method for the thermal history similar to the one observed, we ran a
simulation of a ``realistic'' LCDM model with $\Omega_0=0.3$,
$\Omega_\Lambda=0.7$, $\Omega_b=0.04$, $h=0.7$, $n=1$, $\sigma_8=0.91$
and an equation of state shown by the long-dashed line in Fig.\
\ref{fig:fi}, which is sufficiently similar to the observational data.
We follow exactly the same procedure used to analyze the observations
to derive $T_0$ and $\gamma$.

\def\capfis{% 
The evolution $\gamma$ ({\it a\/}) and $T_0$ ({\it b\/}) as a function
of redshift
for the ``realistic'' LCDM model with sudden \GII\ reionization
at $z \sim 3$ and a photoheating rate 4 times the optically
thin approximation ({\it dashed line\/}).}
\placefig{
\begin{figure}
%\epsscale{0.60}
\inserttwofigures{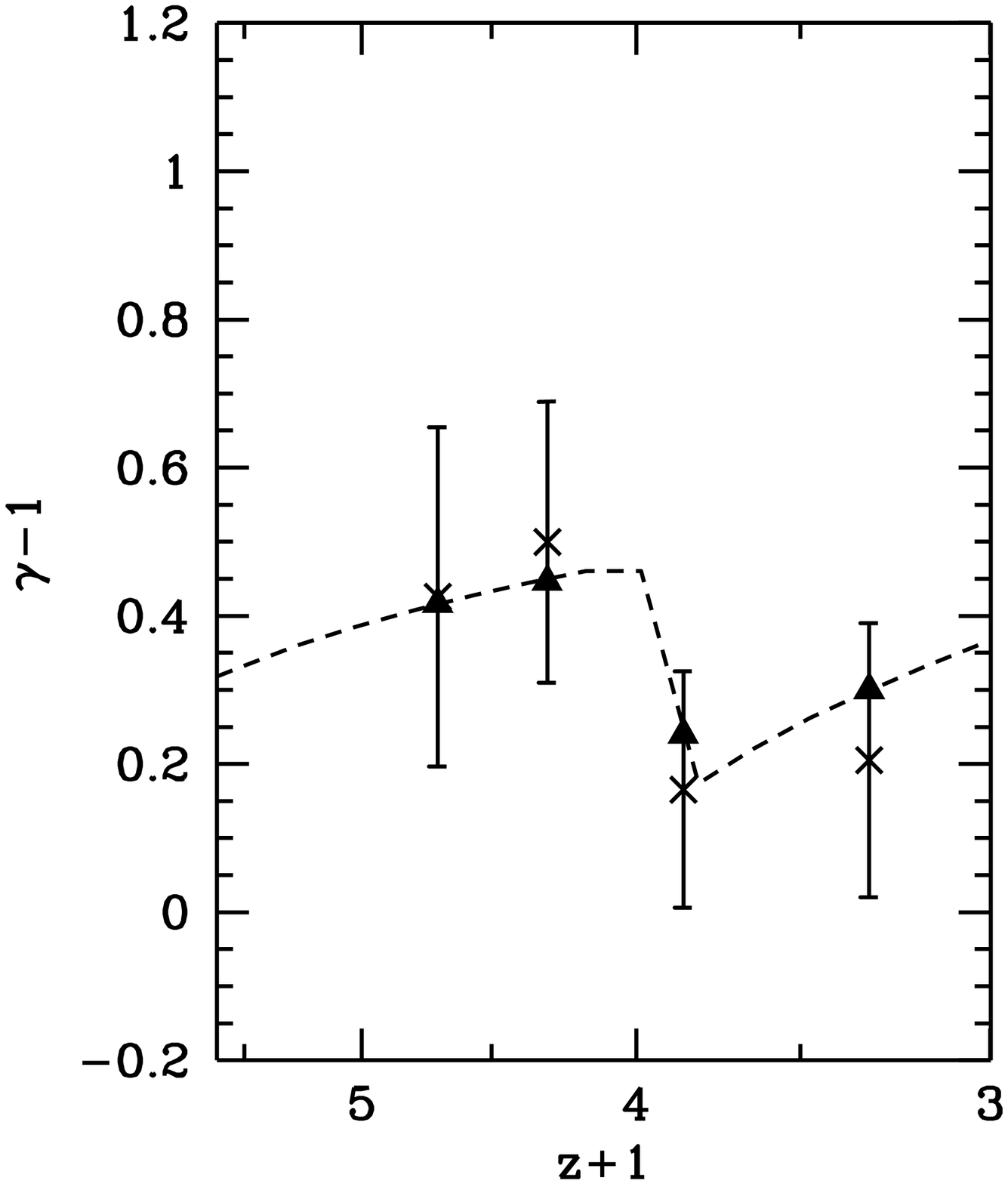}{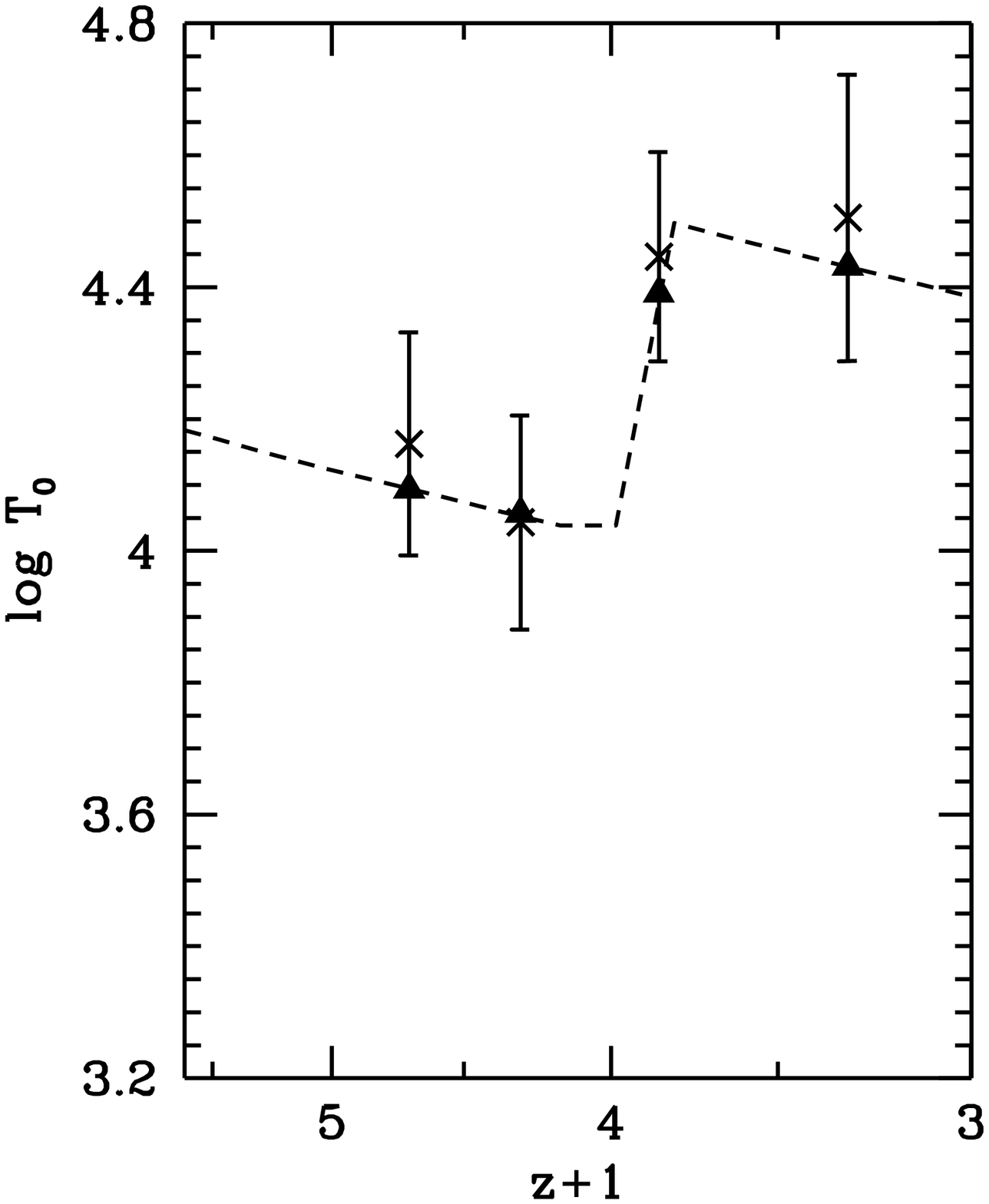}
\caption{\label{fig:fis}\capfis}
\end{figure}
}

Fig.\ \ref{fig:fis} demonstrates that our method works in this case as
well, despite the fact that it was developed, based on the simulations
with the thermal history different from the one observed. In that
figure, we show
%both the 8 models at $z=2.85$ already
%discussed in \S~\ref{ssec:sim} and 
the ``realistic model'' over
mentioned at redshifts $z=3.69, 3.34, 2.85, 2.31$, as well as the
assumed thermal history (dashed line). One can see that our
method reproduces the correct answer within the errors.

Comparing Table~\ref{tab:s2} with Table~\ref{tab:o2}, we note that the
simulated and observed $b-N$ distributions are very similar apart from
the values of the parameter $\varphi_1$. In all simulations but one, 
the parameter
$\varphi_1$, which measures how much the $b$-distribution narrows at
high column densities, is about 3.7. In the observational data, it is
indistinguishable from zero, which implies that the observed
$b$-distribution has the same width for all values of the column
density. In order to understand this difference between the
$b$-distributions we applied our method to a sample of 60 \lya lines
at $z=3$ from a cosmological hydrodynamic simulation of
\markcite{mir96}Miralda-Escud\'e \etal\ (1996), where gas shock heating
is included. For this simulation we found a value $\varphi_1=0.1$,
consistent with the observations. Our interpretation of this result is
that shock heating spreads the $b$-distribution at high column
densities. This does not affect in any way the validity of
our determination of the \eos, but instead can, in principle, be used
to measure the fraction of the gas at a given density that is 
shock-heated.

\section{Summary and Discussion\label{sec:summary}}

We note two main features of Fig.~\ref{fig:fi}: the value of
$\gamma-1$ at $z\sim 3$ is very low and the temperature at $z \sim 3$
is too high. These features are only marginally consistent with simple photoionization models in which \HI\ and \GII\ are photoionized simultaneously at $z\sim5$.
The direct interpretation of this disagreement
is that \HI\ reionization occurs quite late, after $z=5$ or so. This,
however, contradicts the observation of \lya emission lines from quasars
and galaxies at $z>5$ (\markcite{wey99}Weymann \etal\
1999; \markcite{che99}Chen \etal\ 1999), which would be impossible if
the universe was neutral at $z\approxgt 5$ (\markcite{mir98}Miralda-Escud\'e \&
Rees 1998). Another interpretation of our results is that the universe
reheated a second time at $z\sim3$. Reheating will increase
the temperature of the IGM and flatten the effective
equation of state ($\gamma-1$ decreases). One of the most plausible source of
the secondary reheating is \GII\ reionization. Recent observations of the
\GII\ \lya forest also support the conclusion that \GII\ reionization
occurs at $z\sim3$ (\markcite{rei97}Reimers \etal\ 1997;
\markcite{and99}Anderson \etal\ 1999; \markcite{hea99}Heap \etal\
1999). \markcite{ab99}Abel \& Haehnelt (1999) have recently shown that
radiative transfer effects can increase the \GII\ photoheating rate respect
to the optically thin approximation by up to a factor 4. This results in
an increase of the temperature by a factor 1.5-2.5.

The dashed lines in Fig.~\ref{fig:fi} show the effect of a sudden
\GII\ reionization at $z=3$ in the optically thin approximation (the
short-dashed line) and with the \GII\ photoheating rate quadrupled
(the long-dashed line).  The model shown by the dot-dashed line has
the quadrupled \GII\ heating rate and a somewhat gradual \GII\
reionization. Despite being so simple-minded, the latter model
provides a remarkably good fit to the observed thermal history of the
IGM.

Other possible sources of heating in the IGM are photoelectric dust
heating (\markcite{shc99}Nath, Sethi, \& Shchekinov 1999) and Compton
heating by hard X-rays. Recent results (\markcite{mad99}Madau \&
Efstathiou 1999) indicate that the second could be an important
heating source at redshifts larger than $z \sim 10$ but photoheating
is dominant at lower redshifts. Photoelectric heating of dust grains
could be a minor source of heat but some recent observations
show anly a small amount of dust in the high-$z$ \lya forest
(\markcite{out99}Outram \etal\ 1999).

We would like to emphasize here that, while our error bars on the
thermal history of the IGM are quite large, a substantial improvement
can be obtained with a larger data set and extending the completeness
of the sample to low-$N$ lines. The error bars are still dominated by
the statistical errors, as can be seen from comparing
Fig.~\ref{fig:6m} and Fig.~\ref{fig:o1}. With a larger data set, the
errors on $T_0$ and $\gamma$ can be reduced up to a factor of three,
after which the systematic errors become significant. The most
significant systematic error is the lack of knowledge of the intrinsic
shape of the $b-N$ distribution; with the larger data set this error
can also be reduced by introducing a more complicated fit to the
overall distribution. The errors on the parameters $\beta$ and $b_0$
measured with the MLA method are proportional to the square root of
the number of lines of the sample. For a typical simulation with
$10^3$ lines, we have $\Delta \beta \sim 0.06$ and $\Delta b_0 \sim
1.5$ km s$^{-1}$. For the Hu \etal\ (1995) with 636 lines, we have
$\beta \sim 0.08$ and $\Delta b_0 \sim 2$ km s$^{-1}$. The errors on
the \eos parameters are approximately $\Delta (\gamma-1) \la 3.634
\Delta \beta$ and $\Delta T_0/T_0 \ga 2\Delta b_0/b_0$. At redshift
$z\sim 3$ we have six QSOs spectra and the errors are $\Delta
(\gamma-1) \sim 0.1$ and $\Delta T_0/T_0 \sim 9\%$. Thus, more
observations at redshift $z \sim 4$ and $z \sim 2$ are needed to
reduce the current errors. Reducing the errors on the Doppler
parameters and the column densities of the lines are important as
well, in order to obtain an accurate measure of the \eos\ and of the
shape of the $b-N$ distribution. Also, more absorption lines for the
local \lya forest are required in order to put strong constraints on
the evolution of the \eos\ of the IGM at low $z$.

One of the major issues for future work concerns the distribution of
Doppler $b$-values for the low-$z$ \lya absorbers (Penton et
al. 1999). Although we worked here with just 43 absorption lines at
$z < 0.06$, many more lines of sight are currently being analyzed with
HST/STIS spectroscopic data. Thus, theoretical predictions of the
thermal evolution of the IGM at low redshift will soon be testable.
One of the intriguing predictions of many hydrodynamic simulations
(cf., \markcite{cen99}Cen \& Ostriker 1999) is that the relative
fractions of warm (photoionized) absorbers and hot (shocked) shift
considerably between $z = 2$ and $z = 0$. The empirical distribution
of $b$ parameters for Ly$\alpha$ absorbers (from HST) and \lyb
absorbers (from FUSE) should reflect this shift and can provide a test
of these models.

\acknowledgments

We are grateful to Steve Penton for the as yet unpublished data set on
the local \lya forest and to Mark Fardal for the list of published
data on the high-redshift \lya forest. We thank David Tytler, Len
Cowie, and Limin Lu for allowing us to use the published $b-N$
distributions. We also thank Romeel Dav\'e for letting us use his
AUTOVP automated Voigt profile fitting software. This paper was
significantly improved as the result of fruitful conversations with
Andrew Hamilton and Jordi Miralda-Escud\'e.  This work was supported
in part by grants from NASA (NAG5-7262) and NSF
(AST96-17073). Calculations were performed the NCSA Origin2000 array
under the grant AST-960015N.

\placefig{\end{document}}

\clearpage

\newcounter{figurecap}
\setcounter{figurecap}{0}

\begin{center}
\bf Figure Captions
\end{center}

\refstepcounter{figurecap}
Fig.\ \thefigurecap---\label{fig:hcbn}\caphcbn

\refstepcounter{figurecap}
Fig.\ \thefigurecap---\label{fig:hcbb}\caphcbb

\refstepcounter{figurecap}
Fig.\ \thefigurecap---\label{fig:d-N}\capd-N

\refstepcounter{figurecap}
Fig.\ \thefigurecap---\label{fig:b-N}\capb-N

\refstepcounter{figurecap}
Fig.\ \thefigurecap---\label{fig:Pb}\capPb

\refstepcounter{figurecap}
Fig.\ \thefigurecap---\label{fig:cor}\capcor

\refstepcounter{figurecap}
Fig.\ \thefigurecap---\label{fig:6m}\capsm

%\refstepcounter{figurecap}
%Fig.\ \thefigurecap---\label{fig:re}\capre

\refstepcounter{figurecap}
Fig.\ \thefigurecap---\label{fig:lu}\caplu

\refstepcounter{figurecap}
Fig.\ \thefigurecap---\label{fig:o1}\capoa

\refstepcounter{figurecap}
Fig.\ \thefigurecap---\label{fig:o2}\capob

\refstepcounter{figurecap}
Fig.\ \thefigurecap---\label{fig:9r}\cap9r

\refstepcounter{figurecap}
Fig.\ \thefigurecap---\label{fig:fi}\capfi

\refstepcounter{figurecap}
Fig.\ \thefigurecap---\label{fig:fis}\capfis

\clearpage

\tabone
\clearpage

\tabtwo
\clearpage

\tabthree
\clearpage

\tabfour
\clearpage

\tabfive
\clearpage

\tabsix
\clearpage

\tabseven


\begin{references}

\reference{ab99}
Abel, T., \& Haehnelt, M. G. 1999, \apj, 520, L13

\reference{And98}
Anderson, S.\ F., Hogan, C.\ J., Williams, B.\ F., \& Carswell, R.\ F.
1998, \aj, 117, 56

\reference{br99}
Bryan, G. L., \& Machacek, M. 1999, \apj\ submitted (astro-ph/9906459)

\reference{bea99}
Bryan, G. L., Machacek, M., Anninos, P., \& Norman, M.\ L. 1999, \apj, 517, 13

%\reference{bar86}
%Bardeen, J. M., Bond, J. R., Kaiser, N., Szalay, A. S. 1986, \apj, 304, 15
\reference{cen99}
Cen, R., \& Ostriker, J. P. 1999, \apj, 519, L109

\reference{che99}
Chen, H.-W., Lanzetta, K. M., \& Pascarelle, S. 1999, \nat, 398, 586

\reference{cea98}
Croft, R.\ A.\ C., Weinberg, D.\ H., Pettini, M., Hernquist, L., \&
Katz, N. 1999, \apj, 520, 1

\reference{dav97}
Dav\'e, R., Hernquist, L., Weinberg, D. H., \& Katz, N. 1997, \apj, 477, 21
%\reference{dav96}
%Davidsen, A.\ F., Kriss, G.\ A., Zheng, W. 1996, Nature, 380, 47

\reference{da99}
Dav\'e, R., Hernquist, L., Katz, N., \& Weinberg, D. H. 1999, \apj, 511, 521

\reference{efs88}
Efstathiou, G., Ellis, R. S., \& Peterson, B. A. 1988, \mnras, 232, 431

\reference{el99}
Ellison, S. L., Lewis, G. F., Pettini, M.,
Sargent, W. L. W., Chaffee, F. H., Foltz, C. B., Rauch, M., \&
Irwin, M. J. 1999, PASP, 111, 946

\reference{cea98}
Fardal, M. A., Giroux, M. L., \& Shull, J. M. 1998, \aj, 115, 2206

\reference{fer96}
Ferrara, A., \& Giallongo, E. 1996, \mnras, 282, 1165

\reference{gne98}
Gnedin, N. Y. 1998, \mnras, 299, 392

\reference{gh96}
Gnedin, N. Y., \& Hui, L. 1996, \apj, 472, L73

\reference{gh98}
Gnedin, N. Y., \& Hui, L. 1998, \mnras, 296, 44

\reference{haa96}
Haardt, F., \& Madau, P. 1996, \apj, 461, 20
%\reference{hae97}
%Haehnelt, M. G., Steinmetz, M. 1997, \mnras, ???, ??

\reference{hea99}
Heap, S.\ R., Williger, G.\ M., Smette, A., Hubeny, I., Sahu, M., Jenkins,
E.\ B., Tripp, T.\ M., \& Wikler, J.\ N. 1999, \apj, submitted
(astro-ph/9812429)

\reference{her96}
Hernquist, L., Katz, N., Weinberg, D. H., \& Miralda-Escud\'e, J. 1996,
\apj, 457, L51
%\reference{hog97}
%Hogan, C.\ J., Anderson, S.\ F., Rugers, M.\ H. 1997, \aj, 113, 1495

\reference{hu95}
Hu, E. M., Kim, T-S., Cowie, L. L., Songaila, A., Rauch, M. 1995, \aj, 110, 1526

\reference{hg97}
Hui, L., Gnedin, N. Y. 1997, \mnras, 292, 27

\reference{hgz97}
Hui, L., Gnedin, N. Y., Zhang, Y. 1997, \apj, 486, 599

\reference {hr99}
Hui, L., Rutledge, R. E. 1999, \apj, 517, 541
%\reference{Aea98}
%Jakobsen, P., et al.\ 1994, Nature, 370, 35

\reference{ken61}
Kendall, M. G. \& Stuart, A., 1961. The Advanced Theory of Statistics,
Vol. 2, Griffin \& Griffin, London

\reference {kha97}
Khare, P., Srianand, R., York, D. G., Green, R., Welty, D. 1997, \mnras, 285, 167
%\reference {kim97}
%Kim, T-S., Hu, E. M., Cowie, L. L., Songaila, A., 1997, \aj, 114, 1

\reference {kir97}
Kirkman, D., Tytler, D. 1997, \apj, 484, 672

\reference {lu96}
Lu, L., Sargent, W. L. W., Womble, D. S., \& Takada-Hidai, M. 1996, \apj, 472, 509
%\reference {mac98}
%Machacek, M. E., Bryan, G. L., Anninos, P.,
%Meiksin, A., Norman, M. L. 1998, \mnras, ???, ??

\reference {mad99}
Madau, P., \& Efstathiou, G. 1999, \apj, 517, L9

\reference {mir96}
Miralda-Escud\'e, J., Cen, R., Ostriker, J. P., \& Rauch, M. 1996, \apj, 471, 582

%\reference {mir99}
%Miralda-Escud\'e, J., Haehnelt, M., \& Rees, M. J. 1999, \apj, submitted (astro-ph/9812306)

\reference {mir94}
Miralda-Escud\'e, J., \& Rees, M. J. 1994, \mnras, 266, 343

\reference {she99}
Nath, B. B., Sethi, S. K., \& Shchekinov, Y. 1999, \mnras, 303, 1

\reference {out99}
Outram, P. J., Chaffee, F. H., \& Carswell, R. F. 1999, \mnras,
submitted (astro-ph/9907042)

\reference {pen99}
Penton, S. V., Stocke, J.\ T., \& Shull, J. M. 1999, \apj, submitted

\reference{rei97}
Reimers D., K\"ohler S., Wisotzki L., Groote D.,
Rodriguez-Pascal P., \& Wamsteker W., 1997, \aap, 327, 890

\reference {sch99}
Schaye, J., Theuns, T., Leonard, A., \& Efstathiou, G. 1999, \mnras,
in press (astro-ph/9906271)

\reference {shu99}
Shull, J. M., Penton, S. V., \& Stocke, J.\ T. 1999,
Publ. Astr. Soc. Austr., Vol 16, 95
%\reference {shu99}
%Shull, M. J., Penton, S. V., Stocke, J.\ T. 1999, preprint (astro-ph/9901280)

\reference {shu99}
Shull, J. M., Roberts, D., Giroux, M. L.,
Penton, S. V., \& Fardal, M. A. 1999, \aj, 118, 1450

\reference{ssp96}
Shull, J.\ M., Stocke, J.\ T., \& Penton, S. 1996, \aj, 111, 72

%\reference {the98}
%Theuns, T., Leonard, A., Schaye, J., Efstathiou, G. 1998, \mnras, ???, ??

\reference{tea98}
Theuns, T., Leonard, A., Efstathiou, G., Pearce, F.\ R., \& Thomas, P.\ A.
1999, \mnras, 301, 478

\reference{val99}
Valageas, P., \& Silk, J. 1999, \aap, 347, 1

\reference {wei99}
Weinberg, D. H., Katz, N., \& Hernquist, L. 1998, ASP Conference Series,
Vol. 148, eds. C. E. Woodward, J. M. Shull, H. A. Thronson, 21

\reference {wey99}
Weymann, R. J., Stern, D., Bunker, A., Spinrad, H., Chaffee, F. H.,
Thompson, R. I., \& Storrie-Lombardi, L. J. 1999, \apj, 505, L95

\reference {zan97}
Zhang, Y., Anninos, P., Norman, M. L., \& Meiksin, A. 1997, \apj, 485, 496

\end{references}
\end{document}